\documentclass[twocolumn,prb]{revtex4}
\usepackage{subfigure}
\usepackage{color}
\usepackage{array}
\usepackage{amsmath}
\usepackage{amssymb}
\usepackage{wasysym}
\usepackage{bm}
\usepackage{graphicx}
\usepackage{txfonts}
\usepackage{epsfig}
\usepackage[latin9]{inputenc}
\begin{document}

\title{The almost mobility edge in the almost Mathieu equation}

\author{Yi Zhang, Daniel Bulmash, Akash V. Maharaj, Chao-Ming Jian, and Steven A. Kivelson}
\affiliation{Department of Physics, Stanford University, Stanford,
California 94305, USA}
\date{\today}

\begin{abstract}
Harper's equation (aka  the ``almost Mathieu'' equation) famously
describes the quantum dynamics of an electron on a one dimensional
lattice in the presence of an incommensurate potential with
magnitude $V$ and wave number $Q$. It has been proven that all
states are delocalized if $V$ is less than a critical value $V_c=2t$
and localized if $V> V_c$.  Here, we show that this result (while
correct) is highly misleading, at least in the small $Q$ limit.  In
particular,  for $V<V_c$ there is an abrupt crossover akin to a
mobility edge at an energy $E_c$; states with energy $|E|<E_c$ are
robustly delocalized, but those in the tails of the density of
states, with  $|E|>E_c$, form a set of narrow bands with
exponentially small bandwidths $ \sim t\ \exp[-(2\pi\alpha/Q)]$
(where $\alpha$ is an energy dependent number of order 1) separated
by band-gaps $ \sim t Q$. Thus, the states with $|E|> E_c$ are
``almost localized'' in that they have an exponentially large
effective mass and are easily localized by small perturbations. We
establish this both using exact numerical solution of the problem,
and by exploiting the  well known fact that the same eigenvalue
problem arises in the Hofstadter problem of an electron moving on a
2D lattice in the presence of a magnetic field, $B=Q/2\pi$.   From
the 2D perspective, the almost localized states are simply the
Landau levels  associated with  semiclassical precession around
closed contours of constant quasiparticle energy;  that they are not
truly localized reflects an extremely subtle form of magnetic
breakdown.

\end{abstract}

\maketitle

\section{Introduction}

Quasiperiodic ordered states of matter possess two or more periodic
structures whose periods are incommensurate with each other.
Examples of such materials include quasicrystals and crystals with
incommensurate charge or spin density waves as well as simple
crystals in the presence of a generic uniform magnetic field. The
absence of even discrete translational symmetry fundamentally
distinguishes quasiperiodic from crystalline (periodic) structures.
The absence of quenched randomness makes them conceptually distinct
from disordered systems, although because in both cases Bloch's
theorem does not apply, there is no requirement that the elementary
excitations are delocalized.

As one of the simplest examples in the family, Aubry and Andre
studied in 1980 the following one-dimensional finite-difference
Schrodinger equation with a quasiperiodic potential\cite{aubry1980}
\begin{equation}
t\left(f_{n+1}+f_{n-1}\right)+V\cos\left(Q
n+\theta\right)f_{n}=Ef_{n}\label{eq:ame}
\end{equation}
where $n\in\mathbb{Z}$ labels discrete sites, $t$ is the nearest
neighbor hopping amplitude, $V$ and $Q$ are the amplitude and wave
vector of the incommensurate potential ($Q/2\pi\in\left(0,1\right)$
is irrational), and $-\theta/Q$ defines an ``origin'' of the
incommensurate potential. Without loss of generality, we use the
convention in which $V>0$, $t>0$ and open boundary conditions for a
finite size system. This eigenvalue equation is generally referred
to in the physics literature as Harper's equation or the Aubry-Andre
equation and in the mathematics literature as the ``almost Mathieu
equation'' (an analogy with the continuous Mathieu equation). We
will use the name Harper's equation from here on.

Clearly, this is the eigenvalue equation corresponding to a
one-dimensional tight-binding Hamiltonian in the presence of a
quasiperiodic potential, which in second-quantized representation is
\begin{equation}
H_{1D}=\underset{n}{\sum}\left[
t\left(c_{n+1}^{\dagger}c_{n}+c_{n-1}^{\dagger}c_{n}\right)+V\cos\left(Qn+\theta\right)c_{n}^{\dagger}c_{n}
\right] \label{eq:ametb}
\end{equation}
where $E$ is an energy eigenvalue and $\alpha^\dagger
=\underset{n}{\sum} f_n c^\dagger_n $ is the creation operator for
the corresponding eigenstate. The same eigenvalue equation arises in
the ``Hofstadter problem'' of an electron moving in a
two-dimensional square lattice in the presence of a magnetic field
corresponding to $Q/2\pi$ flux quanta per plaquette.

There is a well-known self-duality in Harper's
equation\cite{aubry1980,sokoloff1985, simon1997} which interchanges
the roles of the kinetic and potential energy terms, $V/2
\leftrightarrow t$. Specifically,  in terms of Fourier transformed variables,
\begin{eqnarray}
f_{n} & = & \frac{\exp(i\phi n)}{\sqrt{L}}\underset{m}{\sum}g_{m}\exp\left(imnQ+i\theta m\right)\nonumber \\
g_{m} & = & \frac{\exp(-i\theta
m)}{\sqrt{L}}\underset{n}{\sum}f_{n}\exp\left(-imnQ-i\phi
n\right)\label{eq:duality}
\end{eqnarray}
where $L$ is the system size. 
It is easy to see that $g_{m}$ satisfies the 
dual
Harper's equation
\begin{equation}
\frac{V}{2}\left(g_{m+1}+g_{m-1}\right)+2t\cos\left(Qm+\varphi\right)g_{m}=Eg_{m}.\label{eq:dualame}
\end{equation}
This equation is self-dual when $V = 2t$.

The spectrum and transport properties of this system have been
extensively studied for over thirty years, in a wide set of contexts
in mathematics and statistical mechanics. In particular, the
solution of the ``Ten Martini Problem''\cite{avila2009,last2005}
proves that the spectrum is a Cantor set for $V\ne 0$ and arbitrary
``incommensurate'' $Q$, {\it i.e.} so long as $Q/2\pi$ is
irrational. There has also been progress towards experimental
realization in cold atom systems\cite{Roati}. It is widely
believed\cite{aubry1980,sokoloff1985,kohmoto1991, simon1997,mitmath}
that for any $Q$ that is incommensurate, there is a {\em single}
``metal-insulator" transition at the self-dual point, $V=2t$, such
that the spectrum is absolutely continuous and all eigenstates are
de-localized for $V<2t$, while the spectrum is pure-point and all
eigenstates are localized for $V>2t$. The same line of reasoning
leads to the conclusion that the spectrum is singularly continuous
and the eigenstates are `critical' for $V=2t$.

In this paper, we revisit Harper's equation from various
perspectives. We show that a conventional  semiclassical analysis of
the associated Hofstadter problem suggests that for a weak
potential, $V< 2t$, and small $Q/2\pi\ll 1$, there exist two
mobility edges at $ E_{c,\pm}=\pm\left| 2t-V \right|$ that separate
the localized states with $E<E_{c,-}$ and $E>E_{c,+}$, from the
delocalized states with $E_{c,-}<E<E_{c,+}$; {\it i.e.}, the states
near the band edges are effectively localized even for a weak
incommensurate potential. This conclusion is apparently verified by
``exact'' numerical studies, even when eigenenergies are computed
with an accuracy exceeding 1 part in $10^{12}$. The resolution of
this apparent contradiction comes from a calculable breakdown of the
semiclassical analysis, and a careful analysis of the numerics;
while the states outside the apparent mobility edges are, in fact,
delocalized, they form exponentially narrow bands -- essentially
Landau levels -- with correspondingly large effective masses, $m^*
\sim t^{-1}\exp[(2\alpha\pi/Q)]$, where $\alpha$ is a dimensionless
function of $|E-E_c|$ and $V/2t$. While these states are technically
delocalized, for even moderately small $Q/2\pi$ they are so weakly
dispersing that for all physical purposes they behave as if
localized; we refer to these as ``almost localized'' states,
understanding the physics of which is one of the main results of the
paper. Moreover, a similar analysis in the case $V>2t$, still with
$Q/2\pi\ll 1$, reveals that there is a hidden ``almost transition''
characterized by the emergence of a real-space Fermi surface. For
large wave vector $Q/2\pi\sim O(1)$ and relatively small potential
$V$, we find that perturbation theory gives a satisfactory
characterization and results in a series of gaps in the spectrum,
whose sizes and locations are determined by the values of $V/t$ and
$Q$; all states are robustly de-localized.

The rest of the paper is organized as follows. In Sec.
\ref{sec:semiclassical}, we study the physical properties of
Harper's equation in the limit where $Q\ll2\pi$, analytically using
a semiclassical approximation and by numerical methods, which
suggests the existence of two mobility edges. In particular, because
the system is one dimensional, we are able to employ an extremely
efficient recursive method to obtain numerical solutions for
extremely large system sizes, even with lengths in excess of $10^6$
sites, so we have been able to test the validity of all our analytic
arguments with great precision. In Sec. \ref{sec:paradox}, we
consider high-order perturbations and show that it reconciles the
conflict between Sec. \ref{sec:semiclassical} and previous phase
diagram. Especially, we characterize the essential properties of the
states near the band edges and why the conclusions of Sec.
\ref{sec:semiclassical} is physically relevant. In Sec.
\ref{sec:hiddenPT} we discuss the previously overlooked
``transition'' as a function of energy that occurs  when $Q\ll2\pi$
and $V>2t$, even though all the states are robustly localized. In
Sec. \ref{sec:largeQ}, we briefly present our theoretical
understanding and numerical results for the situation in which the
wave vector $Q\sim O(1)$. We conclude and discuss relations with
previous work in Sec. \ref{sec:conclusion}.

\section{Long period incommensurate order, $Q \ll 2\pi$}
\label{sec:semiclassical}

In this section we present semiclassical and numerical studies
concerning the spectral and transport properties of Harper's
equation when the period of the incommensurate potential is large in
units of the lattice constant.

\subsection{Semiclassical theory of a 2D crystal in a magnetic field}
\label{subsec:2DFS}

It is well known that Harper's equation also describes the
Hofstadter problem - a model of a two-dimensional (2D) crystal in
the presence of an \emph{incommensurate} magnetic flux density. To
establish such an equivalence, consider a 2D tight-binding model on
a square lattice with hopping matrix element $t$ along the $\hat{x}$
direction and $V/2$ along the $\hat{y}$ direction and with an
effective magnetic field $\vec B = B\hat z$ where $B=Q/2\pi$ so that
there are $Q/2\pi$ magnetic flux quanta per plaquette. In a gauge
chosen to preserve translation symmetry along the $\hat{y}$
direction, so that the Bloch wave-number $k_y$ is a conserved
quantity, the Hamiltonian 
 is
\begin{equation}
H_{2D}=\underset{n,k_{y}}{\sum}\left[t\left(c_{n+1,k_{y}}^{\dagger}c_{n,k_{y}}+{\rm
H.C.}\right)+V\cos\left(Qn+k_{y}\right)c_{n,k_{y}}^{\dagger}c_{n,k_{y}}\right]\label{eq:2dproblem}
\end{equation}
where $n$ is the lattice distance along the $\hat{x}$ direction.
Thus, in the subspace of fixed $k_y$, with the identification
$\theta \leftrightarrow k_y$ and leaving the $k_y$ index implicit
$c_{n,k_y} \to c_n$, the Hofstadter problem problem is seen to be
identical to Harper's equation in Eq. \ref{eq:ametb}. Moreover, in
the thermodynamic limit, when $Q/2\pi$ is irrational, spatially
averaged physical quantities such as the density of states (DOS) and
the localization length are independent of $k_y$ for the same reason
that they are independent of the origin of coordinates ($\theta$) of
an incommensurate potential. (See Ref. \onlinecite{qoicdw} and
Appendix \ref{apd:thetaind} for more detailed discussions.) Thus,
the summation over $k_{y}$ in Eq. \ref{eq:2dproblem} only produces a
degeneracy factor of $L_{y}$.

Alternatively, we can choose the gauge that preserves translation
symmetry along the $\hat{x}$ direction:
\begin{eqnarray}
H_{2D'}=\underset{m,k_{x}}{\sum}\left[\frac{V}{2}\left(c_{m+1,k_{x}}^{\dagger}c_{m,k_{x}}+{\rm
H.C.}\right)
+2t\cos\left(k_{x}-Qm\right)c_{m,k_{x}}^{\dagger}c_{m,k_{x}} \right]
\end{eqnarray}
where $m$ and $k_{x}$ label the lattice site and momentum along the
$\hat{y}$ and $\hat{x}$ directions, respectively. In this gauge,
$H_{2D'}$ is equivalent to the dual form of Harper's equation in Eq.
\ref{eq:dualame}. The duality is thus identified with two different
gauge choices for the Hofstadter problem.

\begin{table}
\begin{center}
\begin{tabular}{|c|c|c|c|}
\hline
 & Open along $k_{x}$ & Closed & Open along $k_{y}$\tabularnewline
\hline \hline $V>2t$ & $\left|E\right|<V-2t$ & $\left|E\right|>V-2t$
& N.A.\tabularnewline \hline $V<2t$ & N.A. & $\left|E\right|>2t-V$ &
$\left|E\right|<2t-V$\tabularnewline \hline $V=2t$ & $E=0$ &
$E\neq0$ & $E=0$\tabularnewline \hline
\end{tabular}
\protect\caption{Topologies of the constant energy contours. See
Fig. \ref{fig1} for illustrations. Note that for $V=2t$ and $E=0$
the contour is a perfect square and connected along both the $k_{x}$
and $k_{y}$ directions at isolated points $(\pi,0)$ and $(0,\pi)$.}
\label{table1}
\end{center}
\end{table}

\begin{figure}
\subfigure[ ]{ \includegraphics[scale=0.3]{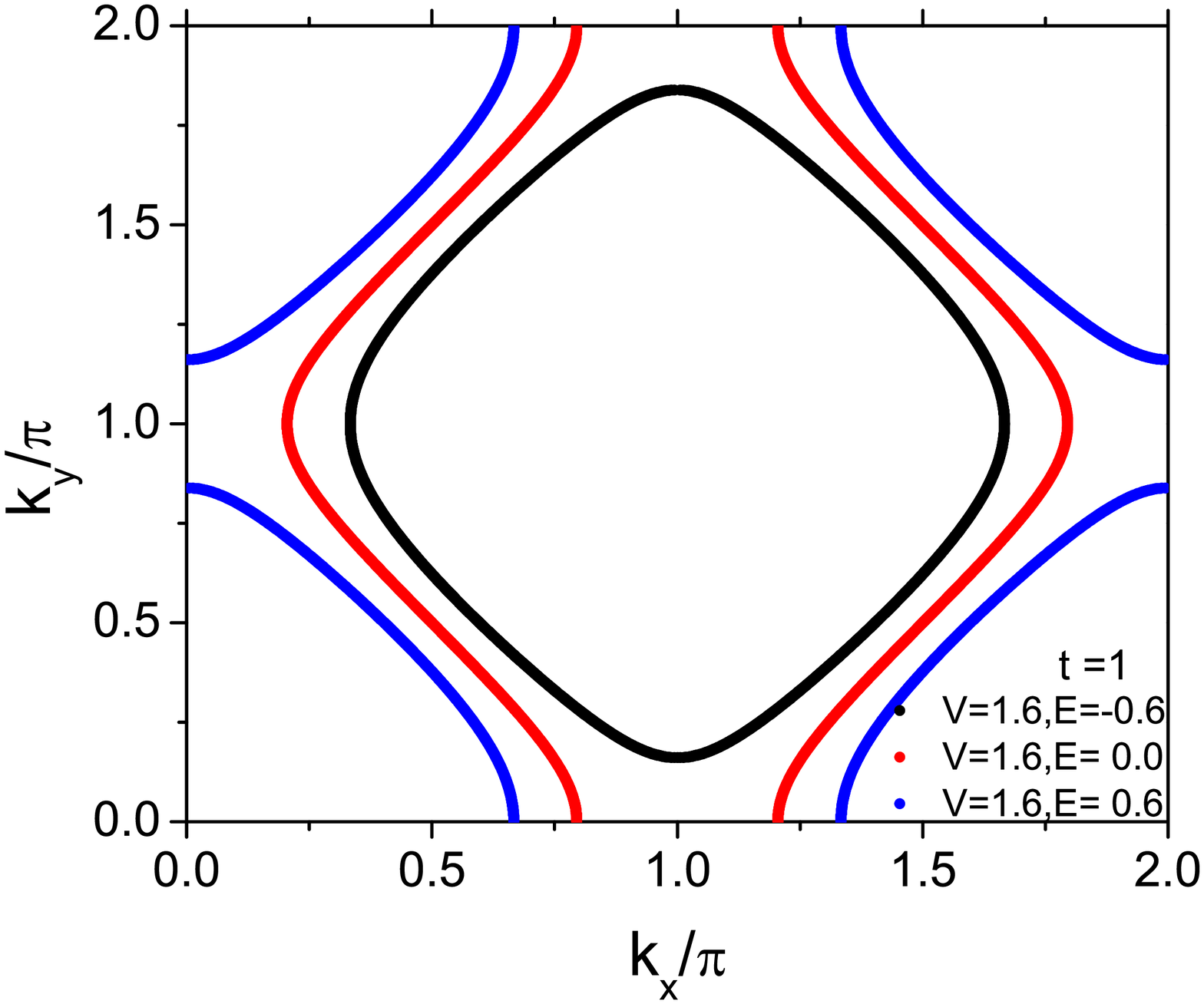}
\label{fig1a}} \subfigure[ ]{
\includegraphics[scale=0.3]{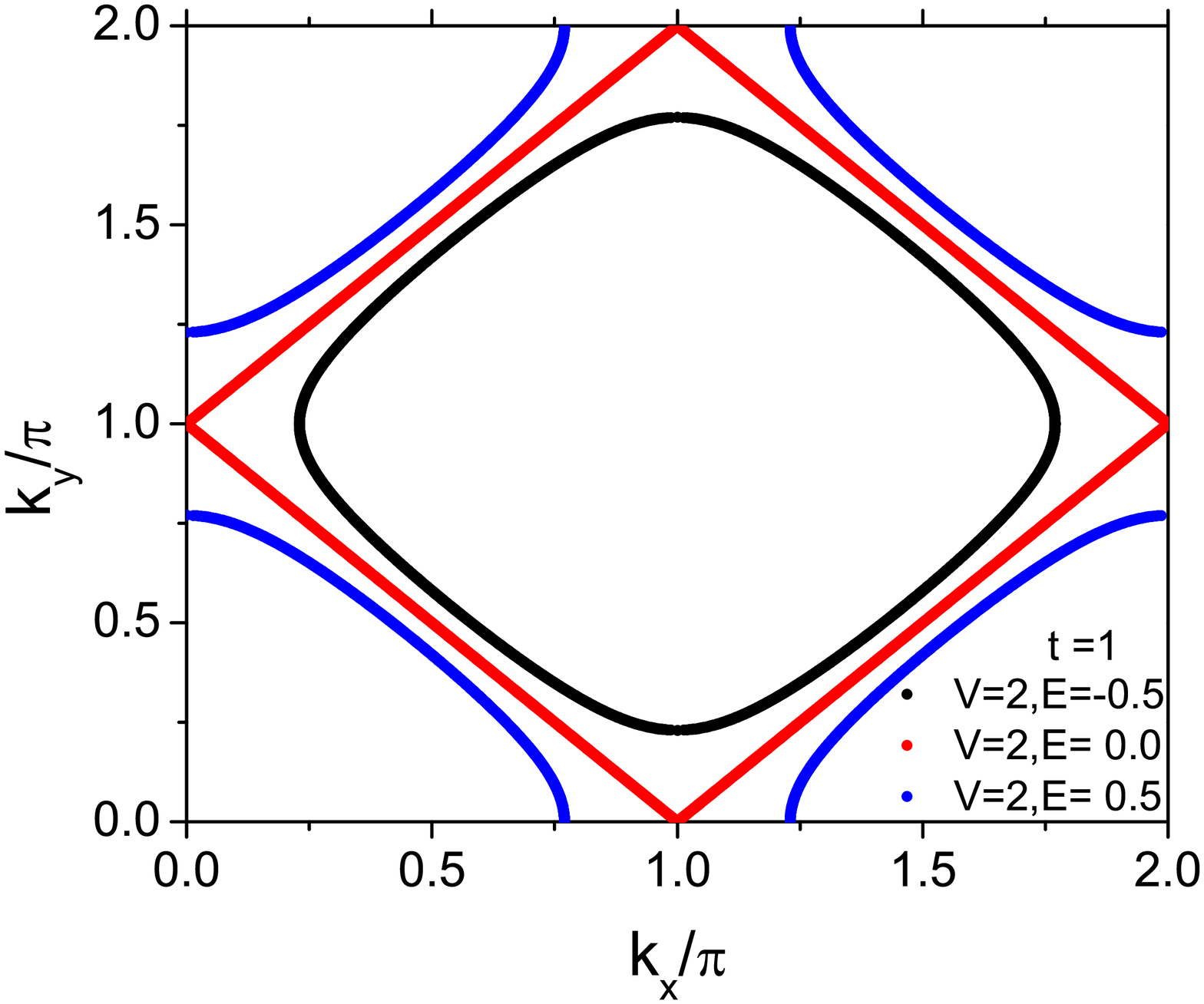} \label{fig1b}} \subfigure[
]{ \includegraphics[scale=0.3]{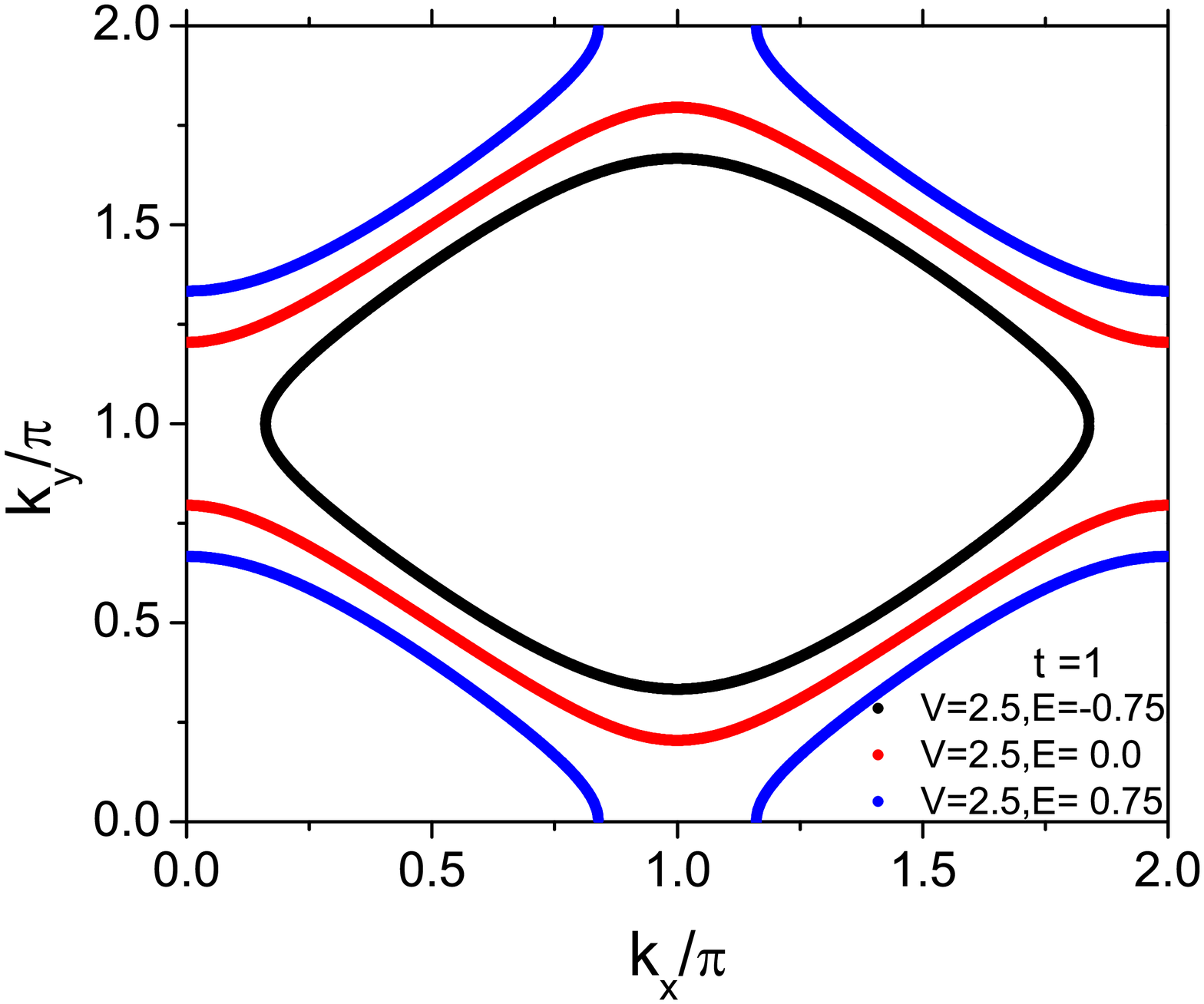} \label{fig1c}}
\caption{Constant energy contours (a) for $V<2t$, (b) at the
self-dual point, $V=2t$, and (c) for $V>2t$. The black curves are at
$E < -\left|2t-V\right|$, the red at $E=0$, and blue at $E >
\left|2t-V\right|$. The black and blue curves are closed contours
while the red one is open in different directions for different $V$
and is at the Lifshitz transition when $V=2t$. Note that $t=1.0$,
$V=2.5$ and $t=0.8$, $V=2$ are equivalent up to an energy rescaling:
$E\rightarrow 0.8 E$.} \label{fig1}
\end{figure}

We can 
study the localization properties of the one-dimensional crystal
with an incommensurate potential by considering whether the
eigenstates of the corresponding two-dimensional crystal in a
magnetic field are localized along the $\hat{x}$ direction, which is
the spatial direction in the original one-dimensional
problem\cite{qoicdw}. When the ``field strength'' $B=Q/ 2\pi \ll 1$,
the dynamics of Bloch electrons is given semiclassically by the
Lorentz force law:
\begin{equation}
\hbar\frac{d\vec{k}}{dt}=-e\vec{v}(\vec{k})\times\vec{B}=\frac{e}{\hbar}\left(\vec{B}\times\frac{dE(\vec{k})}{d\vec{k}}\right)
\end{equation}
where $\vec{v}(\vec{k})=\frac{dE(\vec{k})}{\hbar d\vec{k}}$ is the
electron group velocity. The resulting electron orbits in
two-dimensional momentum space are confined to constant energy
contours determined by the zero-field energy dispersion
$E=\epsilon_{k}=2t\cos k_{x}+V\cos k_{y}$. Since the velocity of an
electron is perpendicular to the constant energy contour, the
semiclassical motion is localized in the $\hat{x}$ direction unless
the contour is open along the $k_{y}$ direction, which occurs when
$V<2t-\left|E\right|$. (See Table \ref{table1} for the parameter
regimes for different topologies of the Fermi surface and Fig.
\ref{fig1} for illustrations.) This seemingly implies that for
potential amplitude $V<2t$ there are two mobility edges at
$E_{c,\pm}= \pm\left(2t-V\right)$; the states are de-localized
between these energies, while all states beyond the mobility edges
are localized. (For $V> 2t$ the same analysis leads to the
conclusion that all states are localized.) If we start from a
pristine one-dimensional crystal and slowly turn on $V$, the states
near the band edges localize first, and as $V$ increases the energy
window of de-localization between the two putative mobility edges
becomes smaller and eventually closes at $V=2t$. This is in sharp
contrast to the expectation that all states are de-localized for
$V<2t$.

The wave vector $Q$ in Harper's equation plays the role of a
magnetic field in the two-dimensional crystal scenario. For
$B=Q/2\pi\ll 1$, the semiclassical results are expected to be
asymptotically exact, and the transition should thus be almost
independent of $Q$. (Exceptions can arise when $V\ll 2t$ or $V\gg
2t$ where the strong anisotropy and curvature of the 2D constant
energy contours invalidates the semiclassical approximation, as will
be discussed further in Sec. \ref{sec:paradox}.) For larger $Q$,
however, the magnetic field can induce tunneling between separate
semiclassical orbits, a phenomenon known as magnetic
breakdown\cite{chambers1966}. In particular, for $Q \sim O(1)$ there
is no reason to trust the semiclassical theory at all, so for the
purposes of the present section, we limit ourselves to small $Q$ and
will only return to consider $Q\sim O(1)$ in Sec. \ref{sec:largeQ}.

To test the validity of our semiclassical arguments, we will now
discuss numerical studies of the localization length and the density
of states (DOS).

\subsection{Numerical results}\label{subsec:numeric}
\label{numerical}

\begin{figure}
\includegraphics[scale=0.3]{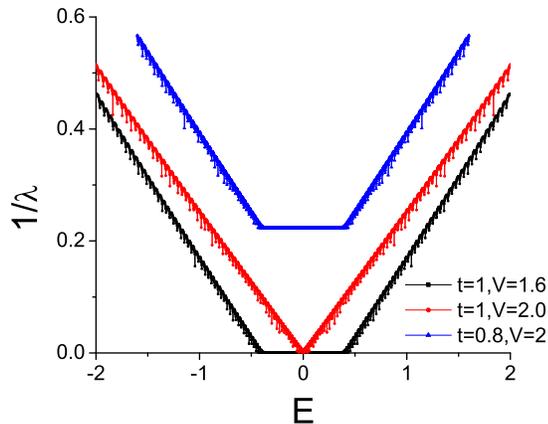}
\caption{The inverse localization length $1/\lambda$ as a function
of energy for three different parameter regimes at $Q=1/31$. The
system size is $L=7.2\times 10^5$. The blue and black curves are
related by the duality transformation in Eq. \ref{eq:duality}
connecting the Hamiltonians in Eqs. \ref{eq:ame} and
\ref{eq:dualame}. The localization length is always finite for the
blue curve, for which $V>2t$ and corresponds to the constant energy
contours in Fig. \ref{fig1c}. However, there is a portion of the
spectrum that is apparently delocalized for the black curve, for
which $V<2t$ corresponding to the constant energy contours in Fig.
\ref{fig1a}. The red curve at the self-dual point $V=2t$ corresponds
to the contours in Fig. \ref{fig1b}.} \label{fig1d}
\end{figure}

Since the one-dimensional Hamiltonian can be written as a
tri-diagonal matrix, numerical calculations are efficient with
recursive methods even for large system sizes\cite{qoicdw}. In
particular, we calculate the two-point Green's function
$G\left(n,n'\right)=\left(E+i\delta-H\right)_{n,n'}^{-1}$. If the
states are localized at energy $E$, then the Green's function should
be exponentially decaying with the distance between the two points
$G\left(n,n'\right)\propto\exp\left(-\left|n-n'\right|/\lambda\right)$
where $\lambda$ is the localization length; otherwise the system is
delocalized and $\lambda\rightarrow\infty$
($1/\lambda\rightarrow0$). In all cases we have carried out
calculations on systems of sizes $L$ large enough that the results
are independent of $L$ - which often means up to sizes $L\sim 10^6$
sites.

We study the localization properties for a representative small
$Q=1/31=2\pi(1/62\pi)\sim 2\pi ( 5.1\times 10^{-3})$ and for various
values of $t$, $V$, and $E$ that lead to different topologies of the
associated two-dimensional constant energy contours as shown in Fig.
\ref{fig1}. In Fig. \ref{fig1d} we display the results of numerical
studies of the inverse localization length as a function of $E$ for
various values of $V/t$. The results are consistent with the
inferences made in the previous section on the basis of the
two-dimensional semiclassical theory: if the energy contour is
closed (black and blue curves in Figs. \ref{fig1a}-\ref{fig1c}) or
open along the $k_{x}$ direction (red in Fig. \ref{fig1c}), the
corresponding states in the one-dimensional (1D) problem are
localized. If the energy contour is open along the $k_{y}$ direction
(red in  Fig. \ref{fig1a}), the 1D states are delocalized.  In
particular, for $V<2t$ but $\left|E\right|>2t-V$ the results
indicate that the states are localized -- consistent with
expectations from semiclassical theory but in conflict with previous
claims\cite{aubry1980,sokoloff1985,kohmoto1991,simon1997,mitmath}.

\begin{figure}
\begin{centering}
\includegraphics[scale=0.3]{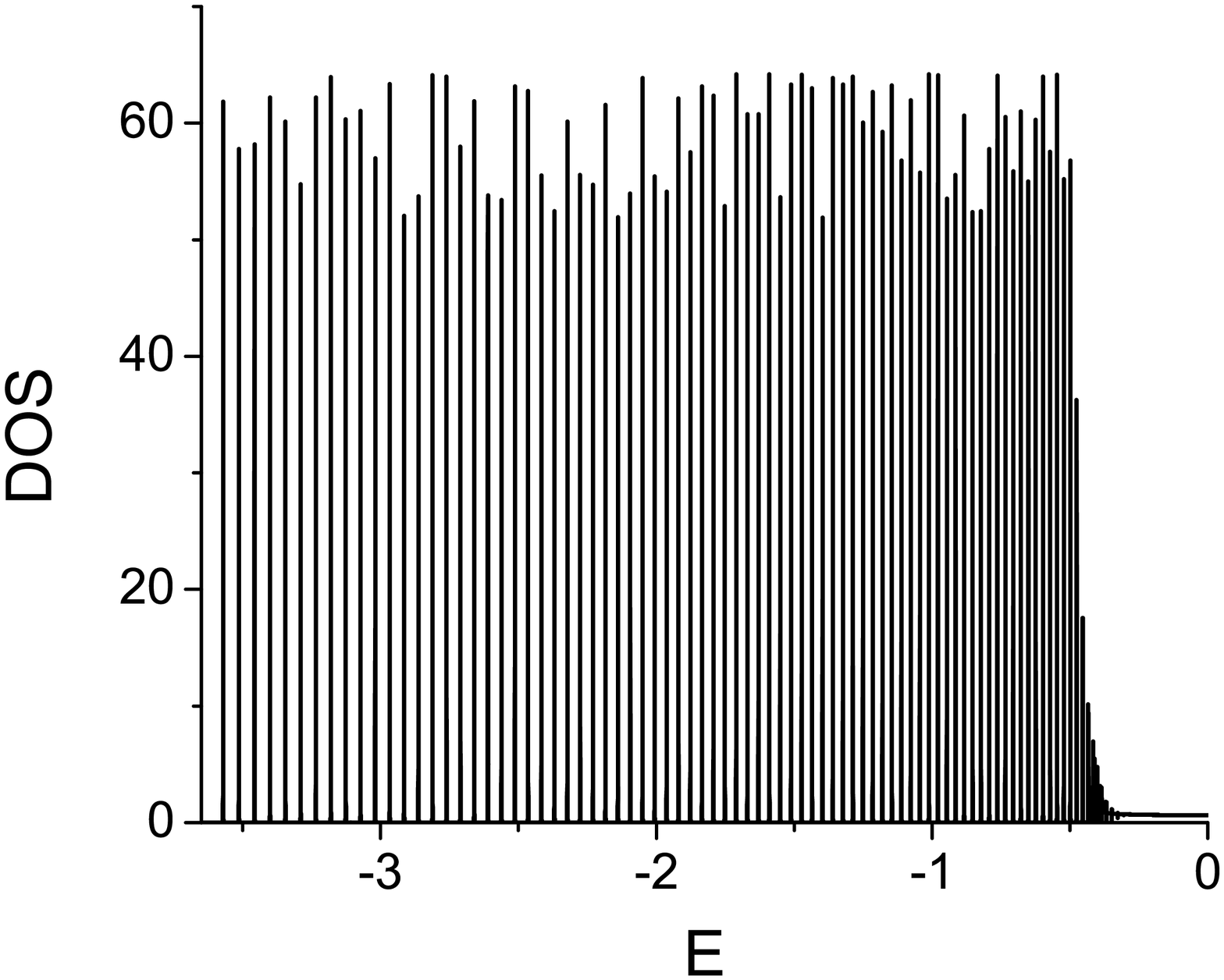}
\includegraphics[scale=0.15]{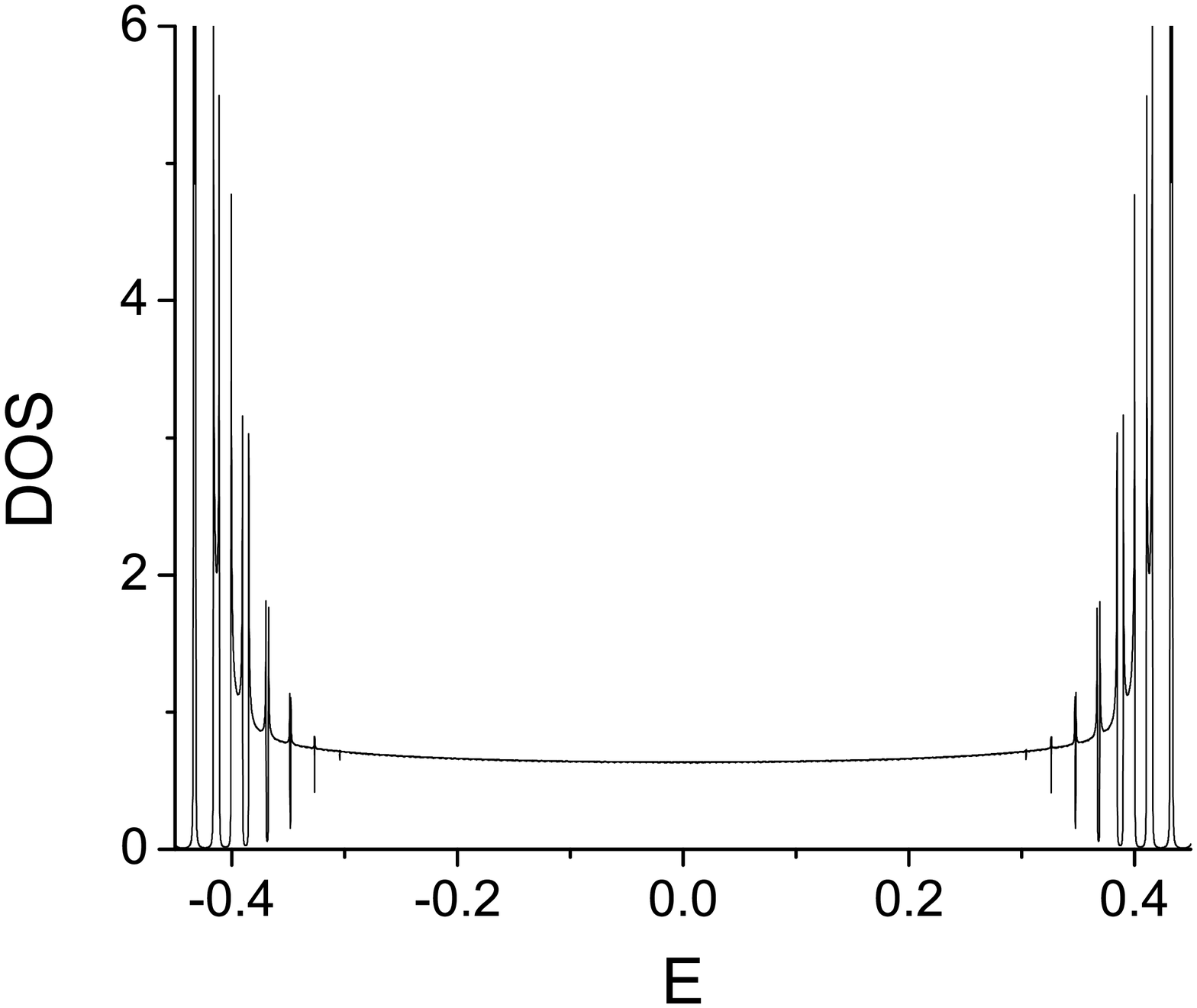}
\includegraphics[scale=0.15]{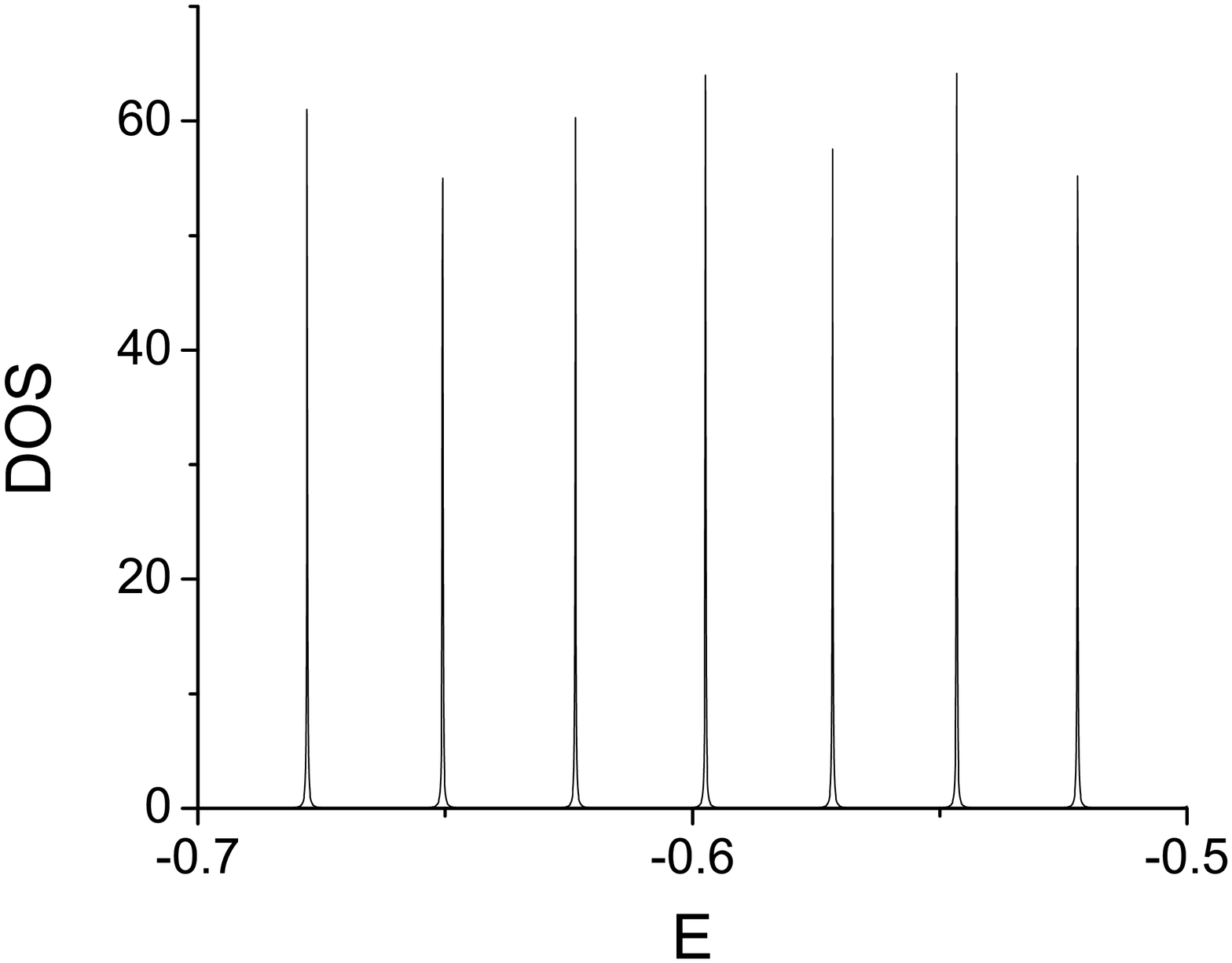}
\protect\caption{The DOS $\rho(E)$ of Harper's equation with
parameters $t=1$, $V=1.6$ and $Q=1/31$. The system size is
$L=7.2\times 10^5$ and $\delta=0.0001$ brings in a small but finite
energy resolution. Drastically different behaviors are observed
above and below $E=\pm0.4$. Note that the spectrum is particle-hole
symmetric therefore only the $E<0$ half is shown. In practice, it is
obtained through the computation for the dual Harper's equation with
$t=0.8$, $V=2$ and $Q=1/31$. The lower panels are regimes between
and just beyond the apparent mobility edges on an expanded scale,
another expansion near the band edge is shown in Fig.
\ref{fig4}.}\label{fig3}
\par\end{centering}
\end{figure}

We have also computed the density of states (DOS) at energy $E$,
defined as $\rho(E)=-\frac{1}{\pi
L}\underset{n}{\sum}\mbox{Im}G(n,n)$, using the same
methods\cite{qoicdw}. Our algorithm is most efficient and accurate
when the system is localized and the Green's function is
exponentially suppressed. However, even if the parameters to be
considered are in the de-localized regime, the spectral properties
are the same as those of the dual Harper's equation given in Eq.
\ref{eq:dualame}, which is necessarily localized\footnote{The
inverse is {\em not} true according to Fig. \ref{fig1d}.}. For this
reason, where the states are delocalized, we always compute the DOS
from the localized dual.

An example of the sort of DOS so obtained is shown in Fig.
\ref{fig3} for $t=1$,$V=1.6$ and $Q=1/31$. Drastically different
behaviors are observed above and below $E=\pm0.4$. For $|E|>0.4$,
there are discrete sharp peaks separated by resolvable gaps while
for $0.4 > |E|$, the DOS appears to be a smoothly varying function
of $E$. The correspondence with the semiclassical treatment of the
2D version is that the discrete peaks correspond to Landau levels
where the constant energy contours (at zero field) are closed. The
continuous spectrum arises where the energy contours are open, and
the apparent mobility edges correspond to the Lifshitz transitions
at which the topology of these contours changes (see Table
\ref{table1}).

\begin{figure}
\begin{centering}
\includegraphics[scale=0.3]{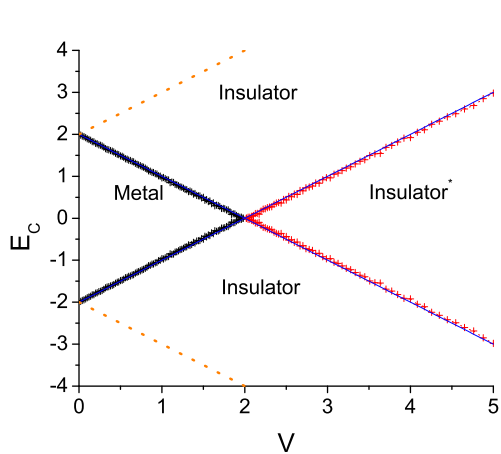}\protect\protect\caption{The phase diagram for Harper's equation
with $t=1$ and $Q=1/31$ obtained from numerical study of
localization length on a system of size $L=7.2\times10^5$. For each
$V<2t$, $E_{c}$ (black) separates the regimes of metal (localization
length $\lambda\rightarrow\infty$) and insulator ({\it i.e.} with
available numerical accuracy, $\lambda$ is finite). For $V>2$, the
critical values of $E_{c}$ (red) are obtained by a duality
transformation of the $V<2t$ cases. Change of DOS behaviors is
observed on both $V<2t$ and $V>2t$ phase boundaries, though on both
sides of the latter phase boundary the system is localized and
insulating. The blue solid lines are $E_{c}=\pm\left(2t-V\right)$ in
accord with Table \ref{table1}. The orange dotted lines are the band
edges at $E=\pm\left(2t+V\right)$. One important caveat is that
there is an implicit energy resolution on the level of machine
precision in our numerical calculations. For infinite precision, the
phase diagram will essentially reduce to a single metal-insulator
transition at $V=2t$, see Sec. \ref{sec:paradox} for more
details.}\label{fig2}
\par\end{centering}
\end{figure}

More generally, Fig. \ref{fig2} is the phase diagram of Harper's
equation with $Q=1/31$ extracted from localization length
calculations. The phase boundaries for $V<2t$ and $V>2t$ are colored
in black and red, respectively, and are fully consistent with
$E_{c}=\pm\left(2t-V\right)$ (blue lines) derived from the
semiclassical treatment of the 2D model, as given in Table
\ref{table1}. On the $V<2t$ side of the phase diagram, 
an apparent metal insulator transition produces qualitative changes
in both the localization length $\lambda$ and the DOS $\rho$. On the
$V>2t$ side, however, while the analogous change in the structure of
the DOS seemingly signifies the existence of a transition (the dual
of the metal-insulator transition) the localization length is finite
on both sides of the transition. We will discuss the nature of this
``insulator-insulator'' transition in Sec. \ref{sec:hiddenPT}.

\subsection{Localization near the band edges from the 2D perspective}
\label{llevels}

It is easy to prove that the spectrum of Harper's equation is
confined to the range of energies $|E| \leq 2t + V$. One important
qualitative point that is less obvious is that the states near the
lower band edge $E\sim-2t-V$ (and the upper band edge, $E \sim
2t+V$) always appear localized, whether or not $V$ is larger than
$2t$. This is most easily understood from the 2D perspective,
$H_{2D}$ in Eq. \ref{eq:2dproblem}. In the absence of an effective
magnetic field ($Q/2\pi =0$), the dispersion near the band-bottom is
accurately treated in the effective mass approximation, which means
that for small but non-zero $Q$, the low energy spectrum is well
approximated as Landau levels. This accounts both for the discrete
peaks in the spectrum and their spatial localization.

\begin{figure}
\begin{centering}
\includegraphics[scale=0.3]{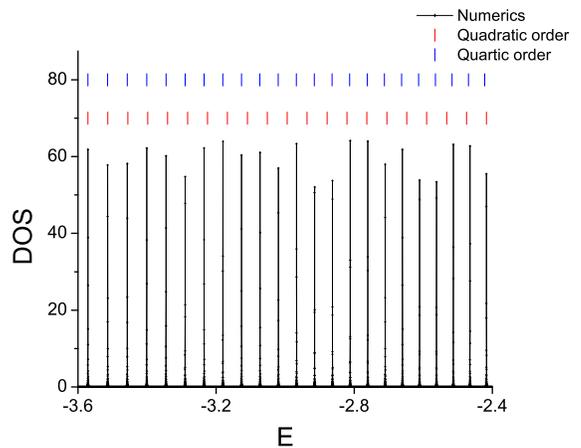}
\protect\caption{The DOS $\rho(E)$ with the same parameters $t=1$,
$V=1.6$, $\delta=0.0001$, and $Q=1/31$ as in Fig. \ref{fig3}, near
the lower band edge but on an expanded scale.. The red and blue
marks on the top label the locations of the discrete Landau levels
from the expansions of the electron dispersion to quadratic order
(free-electron approximation) and quartic order in momentum,
respectively.}\label{fig4}
\par\end{centering}
\end{figure}

To be specific, we expand the zero-field energy dispersion of
$H_{2D}$ to quadratic order of $k$ around the band minimum at
$(\pi,\pi)$:
\begin{equation}
\epsilon_{k}=2t\cos k_{x}+V\cos k_{y}\eqsim-2t-V+t
q_{x}^{2}+\frac{V}{2}q_{y}^{2} + \ldots
\label{harmonic}
\end{equation}
where $(q_{x},q_{y})=(k_{x},k_{y})-(\pi,\pi)$. The effective
magnetic field is $Q/2\pi$ flux quanta per plaquette, and the
corresponding effective cyclotron frequency is
$\omega_{C}=Q\sqrt{2tV}$. Thus, the energy levels near the lower
band edge are quantized in Landau levels with $\epsilon_{n}\approx
-2t-V+\left(n+1/2\right)\omega_C $ where  $n=0,\ 1,\ 2,\ \cdots$. We
show in Fig. \ref{fig4} an enlargement of the DOS near the lower
band edge obtained from numerical solution of $H_{1D}$ with $t=1$,
$V=1.6$, and $Q=1/31$ as in Fig. \ref{fig3}; the corresponding
values of $\epsilon_{n}$ are indicated by the red marks at the top
of the figure.

The consistency is remarkable for the few lowest energy levels,
where the effective mass approximation is highly accurate; at
somewhat higher energy, however, while the energy levels remain
sharp and gapped, the level spacing gradually decreases. However,
the accuracy of the theoretical estimates can be improved by taking
into account the higher order terms in the expansion of the energy
dispersion, $\epsilon_{k}$, e.g. $V\cos
k_{y}\eqsim-V+\frac{V}{2}q_{y}^{2}-\frac{V}{24}q_{y}^{4}+O(q_{y}^{6})$.
The Landau level problem including the quartic terms is discussed in
Appendix \ref{apd:beyondfe}, and the results (obtained with minimal
numerical work\cite{qoicdw}) are shown as the blue marks at the top
of Fig. \ref{fig4}. This accounts quite accurately for the positions
of the peaks in the DOS over the entire range of energies shown in
the figure. The spatial extent of the wave-functions is determined
by the effective magnetic length, $\ell \propto Q^{-1/2}$, which
also determines the relevant range of momenta $\Delta k \sim
1/\ell$. Thus, the validity of the expansion in powers of $k$, and
hence the validity of the Landau level analysis, relies both on the
smallness of $Q$ and on $E$ being close to the band edge, $\omega_C
\ll 4t$ and $|E+2t+V|\ll 4t$. However, {\em at this level of
approximation}, no aspect of the analysis is sensitive to whether or
not $V$ is larger or smaller than $2t$ (as long as the Landau level
spacing $\omega_C$ is much smaller than the original band width,
which requires $(Q/2\pi)^2\ll V/2t \ll (2\pi/Q)^{2}$). We will focus
more carefully on the properties of these band-edge states in Sec.
\ref{sec:paradox}.

Closer to the band center, the fact that there is a periodic lattice
qualitatively affects the Hofstadter problem. As required by
duality, the resulting change from closed to open topology of the
constant energy contours that occurs at $E=\pm E_c$ produces similar
changes in the character of the DOS for $V<2t$ or $V >2t$. For
$V<2t$, there appears to be a continuous DOS for $|E| < E_c$ and the
states are delocalized in the $\hat x$ direction, so in terms of the
properties of the original 1D problem, $E_c$ acts as a mobility edge
in the traditional sense. However, for $V>2t$, from the perspective
of the original 1D problem, the states are robustly localized both
above and below $E_c$; the nature of the crossover that occurs at
$E=\pm E_c$ in this case will be discussed in Sec.
\ref{sec:hiddenPT}.

\section{Resolving the paradox - delocalization
of the states near the band edges\label{sec:paradox}}

As we have shown in Sec. \ref{sec:semiclassical}, the semiclassical
and numerical results disagree qualitatively with the previous
literature concerning both the localization and spectral properties
of Harper's equation for small $Q\ll2\pi$. In this section, we first
briefly summarize the previous arguments that there is a single
energy independent metal-insulator transition at $V=2t$, and then
examine more carefully the properties of the states near the band
edges where the contradiction arises. In particular, we focus on two
perturbations that have been neglected in the semiclassical
approximation and show how they affect the localization problem. We
will see, for $Q\ll2\pi$ and $V<2t$, the states near the band edges
behave `almost localized': even though their wavefunctions are
extended, their bandwidth are exponentially small in $1/Q$,
therefore they are easily localized in the presence of small
perturbation or finite energy resolution.

\subsection{Previous arguments on the localization transition\label{subsec:previous}}

First of all, we briefly summarize the previous
arguments\cite{aubry1980, sokoloff1985}. In the notation of Eq.
\ref{eq:duality}, we define
\begin{equation}
f(x)=\frac{\exp(i\phi x)}{\sqrt{L}}\underset{m}{\sum}g_{m}\exp\left(imQx+i\theta m\right)
\end{equation}
as an extension of $f_{n}$, $n \in \mathbb{Z}$ to $x\in\mathbb{R}$.
Since $m$ takes integer values, for a $g_{m}$ which remains
normalizable in the $L \rightarrow \infty$ limit (that is, $g$ is
localized), $f(x)$ is a Bloch function with a fundamental period of
$2\pi/Q$. Then the wavefunction $f_{n}$, though no longer periodic,
is still extended due to its periodic envelope $f(x)$. Namely, if
the solution to one specific Harper's equation is localized, its
dual is necessarily extended.

We can further use the fact that the inverse localization length is
related to the DOS through the Thouless formula:
\begin{equation}
1/\lambda(E)=\int dE' \ln \left|(E-E')/t
\right|\rho(E')\label{eq:thouless}
\end{equation}
which can be equally applied to the dual system in Eq.
\ref{eq:dualame} with the same DOS $\rho$:
\begin{equation}
1/\lambda'(E)=\int dE' \ln \left|2(E-E')/V \right|\rho(E') \label{eq:dualthouless}
\end{equation}
then $1/\lambda(E)=1/\lambda'(E)+\ln(V/2t)$. This relation holds
well as can be seen in Fig. \ref{fig1d}: the Harper's equations
corresponding to the blue and black curves are dual, and indeed
there is a constant vertical shift of $\ln(V/2t)$ between them.

Let us first consider the case where states in the original system
are localized: its dual is extended so $1/\lambda'(E)=0$, therefore
$1/\lambda(E)=\ln(V/2t)>0$ is constrained to $V>2t$. Similarly, if
the original system is extended $1/\lambda(E)=0$, its dual
$1/\lambda'(E)=\ln(2t/V)>0$ requires $V<2t$. Consequently, all
states are localized at all energy levels $E$ for $V>2t$, and all
states are extended at all energy levels $E$ for $V<2t$.

There is an obvious contradiction between this argument and the
two-dimensional semiclassical problem, where the duality corresponds
to switching the roles of $\hat{x}$ and $\hat{y}$. For open constant
energy contours (red and blue curves in Figs. \ref{fig1a} and
\ref{fig1c}), the semiclassical theory does suggest that the system
is either localized along $\hat{x}$ and de-localized along
$\hat{y}$, or localized along $\hat{y}$ and de-localized along
$\hat{x}$. However, there is a third possibility that the constant
energy contour can be closed (black curves in Figs. \ref{fig1a} and
\ref{fig1c}), where the states are physically localized along both
the $\hat{x}$ and $\hat{y}$ directions. More specifically, while the
previous claim that all states are localized for $V>2t$ and the
states between the two apparent mobility edges, i.e. those with
$E_c=\pm(2t-V)$ are de-localized for $V<2t$ is consistent with the
results in Sec. \ref{sec:semiclassical}, both the semiclassical
theory and the numerics seemingly imply that the states beyond the
mobility edges at $\pm E\in(2t-V,2t+V)$ are localized even for
$V<2t$.

\subsection{Localization near the band
edges from a real-space perspective}\label{subsec:1dperturb}

In Sec. \ref{llevels}, and more generally in Sec. \ref{subsec:2DFS},
we found that the states near the lower band edge form effective
Landau levels; at this level of approximation, the states in a
Landau-level are exactly degenerate, and thus it is possible to
construct eigenstates that are either localized or delocalized.  In
this circumstance, even parametrically small terms that are omitted
can, in principle, play a qualitative role in resolving this
degeneracy. In this subsection we consider the states near the lower
band edge from a 1D real-space perspective, treating $Q\ll2\pi$ as a
small parameter. In particular, we show that exponentially small
terms, $\sim \exp[-(2\alpha \pi/Q)]$  lift this degeneracy in such a
way that the band-edge states are delocalized for $V<2t$ and
localized for $V>2t$.  Note that terms that produce these behaviors
are so small for even moderate values of $2\pi/Q$ that they are
entirely unobservable in the numerical studies of Sec.
\ref{numerical}, despite their being carried out to less than 1 part
in $10^{12}$ uncertainty! Conversely, since it is difficult to
imagine circumstances in which the energy uncertainty (due to
thermal broadening if nothing else) is anywhere near this small,
this observation carries with it the implication that {\it for all
plausible physical purposes, these states are effectively
localized.} It is in this sense that they are ``almost localized.''

Consider the states within a quantized energy level $\epsilon_n$ and
the corresponding Hilbert subspace consisting of a single localized
state per potential well consisting of $~2\pi/Q$ sites. There are
two perturbations that are exponentially small in $1/Q$ and
implicitly neglected in the semiclassical theory in Sec.
\ref{sec:semiclassical} -- the tunneling between the potential wells
$t_{\rm eff}$ and the local energy variation between the potential
wells $V_{\rm eff}$, which turn out to be the key to resolving the
conflicting conclusions. The way this occurs can be most easily seen
from a simple variational analysis of the band-bottom states in the
limit $Q \ll 1$.

\emph{Local energy difference between wells:} To begin with, we
construct a variational state centered near a single minimum of $V$
of the gaussian form
\begin{equation}
f_n = {\cal N} e^{i\pi n} \exp[-(n-a)^2/2\ell^2]
\end{equation}
where ${\cal N}$ is the normalization constant, while the center of
localization, $a$, and the localization length, $\ell$, are treated
as variational parameters, although we will assume (and then
confirm) that $2\pi/Q\gg \ell \gg 1$. The variational energy in this
state is easily seen to be (using the Poisson summation formula)
\begin{equation}
\langle H \rangle = V\cos(Qa)e^{-Q^2\ell^2/4} - 2te^{-1/4\ell^2} +\delta E
\end{equation}
where $\delta E$ is all terms of order $e^{-\pi^2m^2\ell^2}$ with
integer $m \geq 1$. Clearly, the leading order terms (ignoring
$\delta E$) simply reproduce the results of  the harmonic
approximation in Eq. \ref{harmonic}.  The result is $a =
\pi(1+2m)/Q$, $\ell = (2t/V)^{1/4}\sqrt{1/Q}$ giving a variational
energy  $\epsilon_0\approx -2t-V+ Q\sqrt{2tV}/2$ that is independent
of $m$, {\it i.e.} the minimum of the potential in which it is
localized.

This result is not exact. The leading contribution to its
corrections can be readily seen by looking at the leading $m$
dependent term in $\delta E$: $\delta E =
-2Ve^{-\pi^2\ell^2}\cos(2\pi a) + \ldots $ where $\ldots$ refers to
other terms of the same order which are independent of $a$, as well
as higher order terms in powers of $e^{-\pi^2\ell^2}$.  Thus, we see
that there is an exponentially small difference in energy
\begin{equation}
V_{\rm eff} \sim V \exp[-\pi^2\ell^2]
\end{equation}
between a state localized near $a=\pi/Q$ and the neighboring well at
$a=3\pi/Q$.

\emph{Tunneling between the potential wells:}  Using the same
variational wavefunctions, we can readily estimate the tunneling
matrix element between two neighboring states by simply evaluating
the matrix element of the tunneling term in the Hamiltonian between
localized states centered at $a=\pi/Q$ and $a=3\pi/Q$.  The result
is
\begin{equation}
t_{\rm eff} \sim t \exp[-\pi^2/Q^2\ell^2]. \label{eq:teff}
\end{equation}
this is qualitatively consistent with the more careful analysis
included in Appendix \ref{apd:functionT}.

From this simple variational treatment, we conclude that $V_{\rm
eff} \gg t_{\rm eff}$ ({\it i.e.} the states are localized) so long
as $Q^2\ell^4 <1$, and conversely, that $V_{\rm eff} \ll t_{\rm
eff}$ if $Q^2\ell^4 >1$.  Miraculously (and probably accidentally),
since this  variational approach yields $Q^2\ell^4=2t/V$, it
reproduces the exact criterion for localization.  (Note that in
terms of parametric dependence on $Q$, $\ell^2 \sim Q^{-1/2}$, which
implies that both $\ell^{-2}$ and $Q^2\ell^2$ are small compared to
1 when $Q\ll1$, as promised.)

\subsection{Localization near the band edge from the perspective of magnetic breakdown}
To locate the localization transition more accurately, we consider
related corrections to the 2D semiclassical theory from Sec.
\ref{subsec:2DFS}. So long as the effective magnetic field,
$Q/2\pi$, is non-zero, the semiclassical theory is not exact, and in
particular there is always a non-zero amplitude for magnetic
breakdown across the Brillouin zone boundaries. While the amplitude
for such processes is exponentially small, they are exactly the
terms that determine whether the states in a given Landau level are
localized or not.

Magnetic breakdown along the $\hat y$ and $\hat{x}$ directions
produce essentially different outcomes: the former results in an
open constant energy contour along the $\hat y$ direction and hence
de-localization along the $\hat x$ direction, and thus is associated
with $t_{\rm eff}$; the latter results in an open constant energy
contour along the $\hat x$ direction and hence is associated with
local energy differences in potential wells, $V_{\rm eff}$.
Manifestly, when $V>2t$, the constant energy contour near the band
edge is anisotropic and elongated along the $\hat{x}$ direction, and
conversely for $V<2t$. (See Fig. \ref{fig1}.)  Therefore magnetic
breakdown occurs predominantly along the $\hat{x}$, so that the
states are localized suggesting $V_{\rm eff}>2t_{\rm eff}$, or
$\hat{y}$ direction so that the states are de-localized suggesting
$V_{\rm eff}<2t_{\rm eff}$, depending on whether $V/2t$ is greater
or less than 1. Indeed, the constant energy contours are symmetric
under $\pi/2$ rotation only at the self-dual point of $V=2t$, which
means that only here is $V_{\rm eff}/2t_{\rm eff}=1$. A further
implication of this is that the spectrum forms a Cantor set, since
for each Landau band of the original problem, {\it i.e.} for each
value of the energy level $\epsilon_n$, the effective Hamiltonian
confined to the space of states spanned by this Landau level defines
a new version of Harper's equation, but with much smaller energy
scale and longer length scale. These results are indeed consistent
with previous
conclusions\cite{aubry1980,sokoloff1985,kohmoto1991,simon1997,mitmath}.

\begin{figure}
\begin{centering}
\includegraphics[scale=0.3]{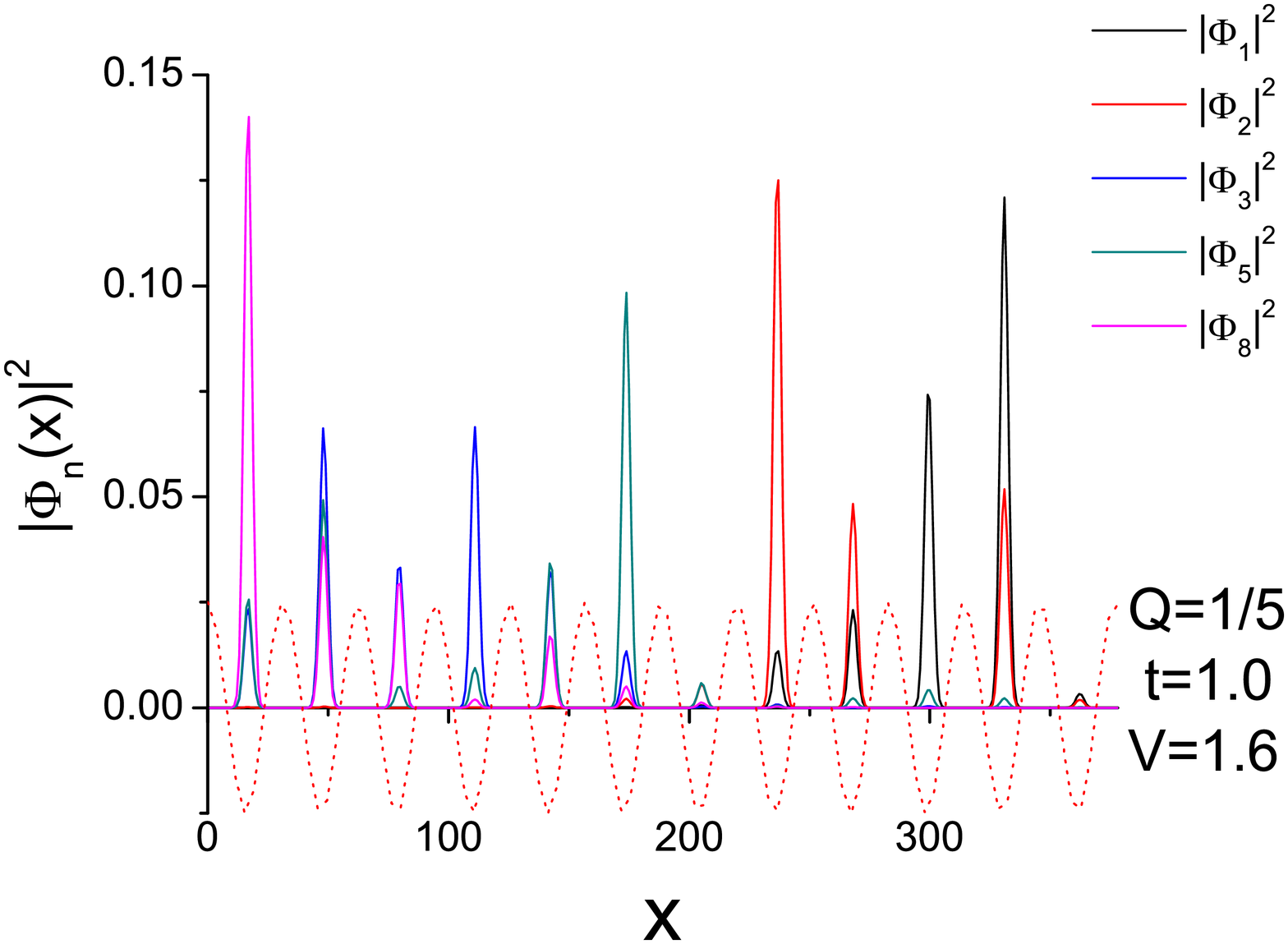}
\includegraphics[scale=0.3]{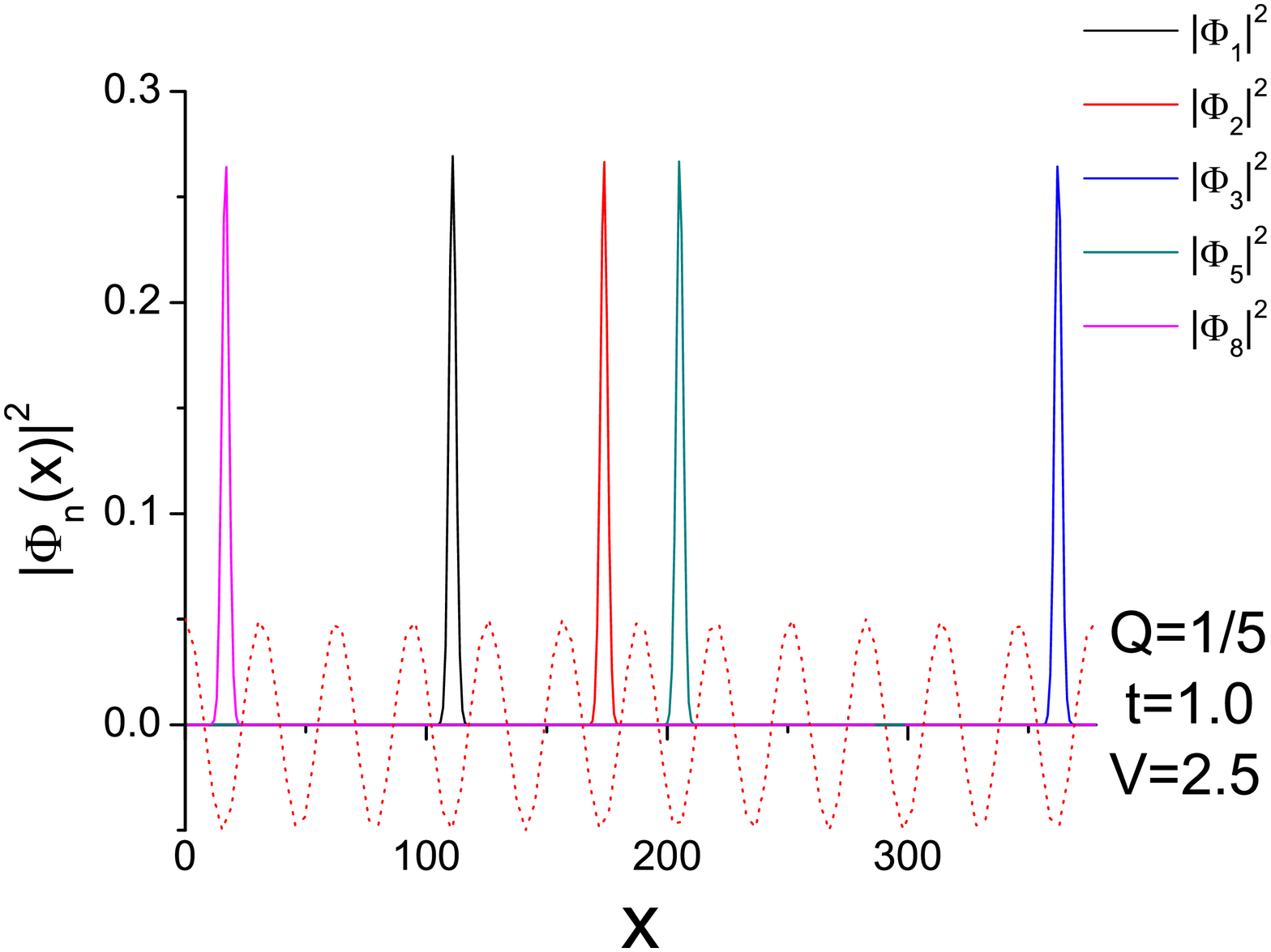}
\protect\caption{The probability amplitude of lowest eigenstates of
Harper's equation for $Q=1/5$, $t=1$, $V=1.6$ (upper panel) and
$V=2.5$ (lower panel). The results are obtained through exact
diagonalization on system size $L=377\eqsim12\left(2\pi/Q\right)$
with almost $12$ complete potential wells. The dotted red curves are
illustrations of the $V\cos\left(Qx\right)$ potential profile (not
to scale).}\label{fig12}
\par\end{centering}
\end{figure}

\subsection{Numerical tests for ``moderate'' $Q/2\pi$}
To test this perspective, we numerically exact diagonalize Harper's
equation Hamiltonian for a representative value of $Q=1/5$, which is
still small compared to $2\pi$, but not so small that exponentially
small effects are beyond the realm of numerical studies (even by
setting $\delta=0$). We take $L=377\eqsim12\left(2\pi/Q\right)$, the
probability distribution of some lowest eigenstates for $t=1$ and
$V=1.6$ (upper panel) or $V=2.5$ (lower panel) are shown in Fig.
\ref{fig12}. We can clearly see that the states near the lower band
edge are localized for $V=2.5t>2t$ but clearly extended for
$V=1.6t<2t$ -- once we can resolve the exponentially small
perturbations.

We note that similar arguments can be useful for the identification
of the metal-insulator transition and mobility edges in the
generalized versions of Harper's equations such as the examples
considered in Ref. \onlinecite{gaa}, where one can simply compare
the quantum tunneling amplitudes across the $\hat x$ and $\hat y$
directions in the equivalent two-dimensional Hofstadter's problem.
Further details and examples are included in Appendix \ref{apd:gaa}.

\section{Hidden crossover for $V>2t$ and $Q\ll2\pi$: a real-space Fermi surface \label{sec:hiddenPT}}

In the previous sections, we have discussed the effective
metal-insulator transition for Harper's equation with $V<2t$ and
$Q\ll2\pi$, where drastically different behaviors of the spectral
and transport properties are observed in the physical limit on the
two sides of the apparent mobility edges $\pm E_{c}=2t-V$. In the
current section, let us turn to the related `insulator-insulator'
transition at $\pm E_{c}=V-2t$ in the duality transformed systems
for $V>2t$ and $Q\ll2\pi$, see Fig. \ref{fig2}. All states are
localized for $V>2t$, yet the behavior of the DOS still shows a
sharp change in character at $E_{c}$, see Fig. \ref{fig3}. What is
the nature of this transition?

First of all, the Pauli exclusion principle forbids the electron
density per site exceeding one. As we have shown in Sec.
\ref{sec:semiclassical}, near the band edges the electron
eigenstates of Harper's equation localized in each potential well
resembles that of a harmonic oscillator with their electron density
mostly concentrated around the potential minima. Heuristically, for
a deep potential well, the electron density there will reach one
electron per site after filling a finite number of quantized levels.
For $Q\ll2\pi$ the potential around the minima is slow varying,
therefore the nearby lattice sites have a band width approximately
equal to $W=4t$ -- the band width of an infinite 1D chain with
$V=0$. From the lower band edge $E=-2t-V$, these sites will reach
maximum electron density at $E=2t-V$ consistent with $E_{c}$ if
$V>2t$.

\begin{figure}
\begin{centering}
\includegraphics[scale=0.3]{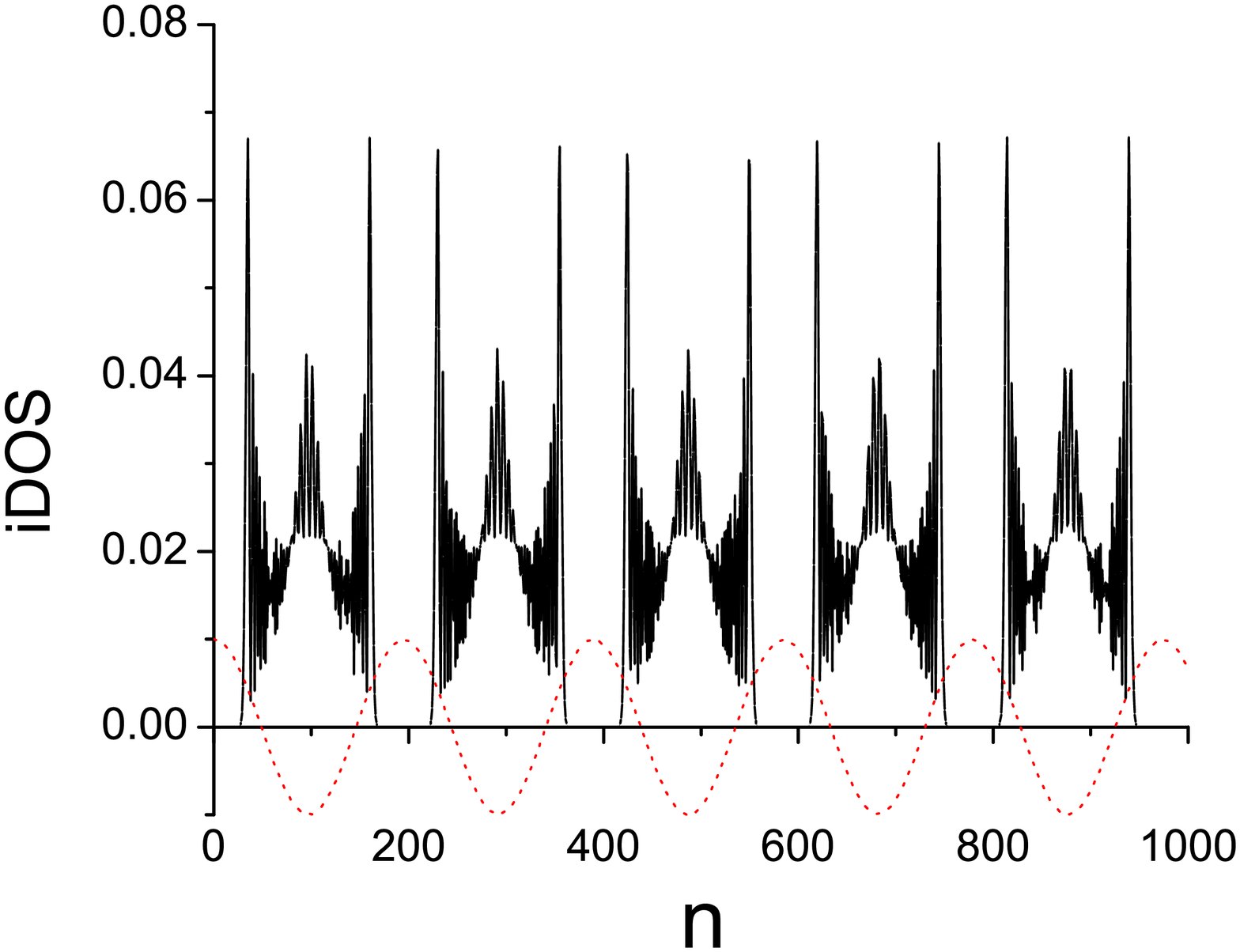}
\includegraphics[scale=0.3]{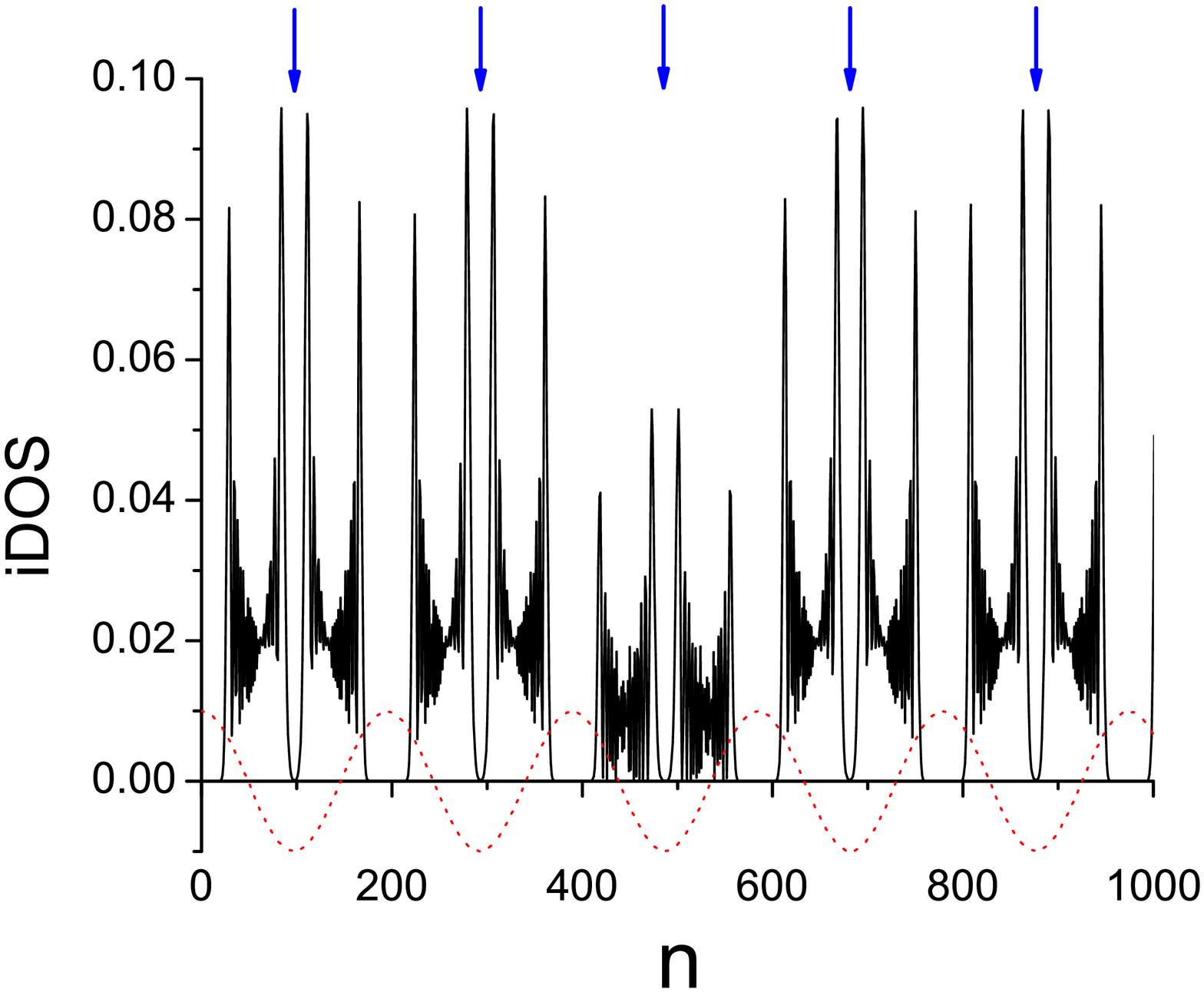}\protect
\protect\caption{The local electron integrated DOS of Harper's
equation $\int_{E_{min}}^{E_{max}}dE \left|f_n(E)\right|^2$ with
$t=1$, $V=2.5$ and $Q=1/31$ for states within the energy range
$-0.8<E<-0.7$ (upper panel) and $-0.4<E<-0.3$ (lower panel),
respectively. The red dotted curve is an illustration of the
incommensurate periodic potential (not to scale). The blue arrows in
the lower panel mark additional dips appearing at the potential
minima.}\label{fig5}
\par\end{centering}
\end{figure}

\begin{figure}
\begin{centering}
\includegraphics[scale=0.3]{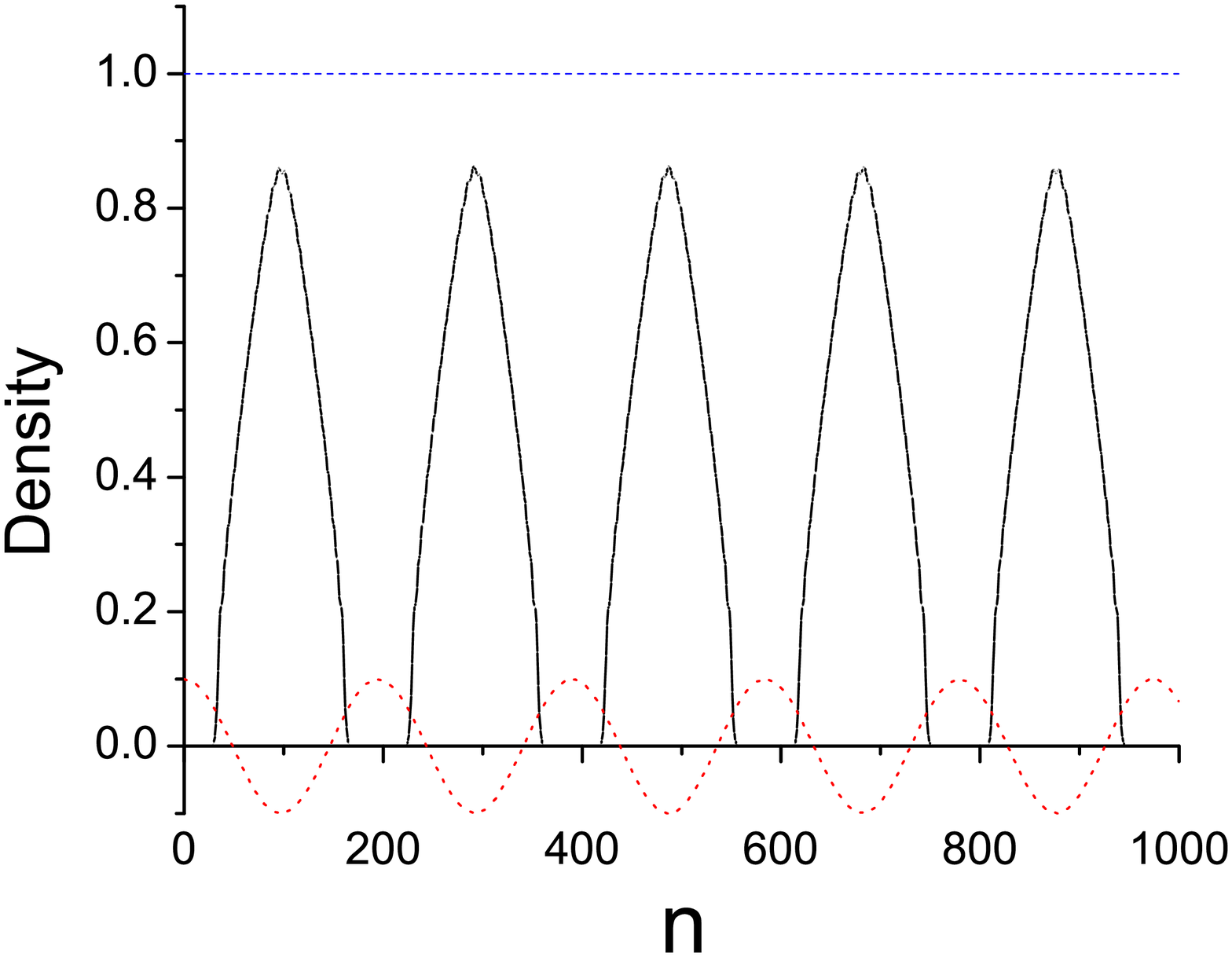}
\includegraphics[scale=0.3]{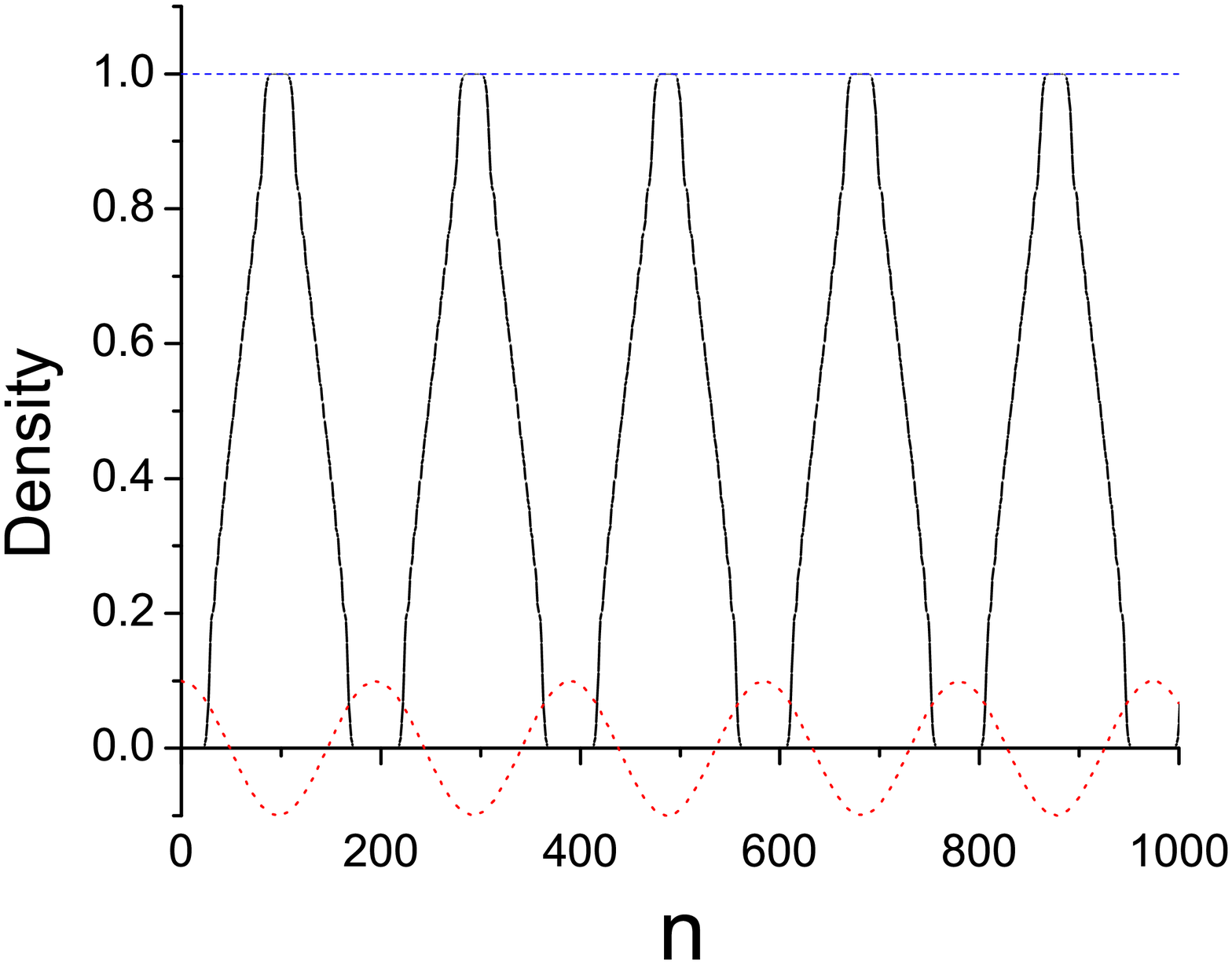}\protect
\protect\caption{The real-space electron density of Harper's
equation $\int_{-2t-V}^{E_F}dE \left|f_n(E)\right|^2$ with $t=1$,
$V=2.5$ and $Q=1/31$ for all states below the Fermi energy
$E_F=-0.7$ (upper panel) and $E_F=-0.3$ (lower panel), respectively.
The blue dashed line indicates the maximum electron density allowed
- one electron per site. The red dotted curve is an illustration of
the incommensurate periodic potential (not to scale). }\label{fig9}
\par\end{centering}
\end{figure}

For illustration, we calculate the local electron integrated DOS of
Harper's equation within an energy range $\int_{E_{min}}^{E_{max}}dE
\left|f_n(E)\right|^2$ with $t=1$, $V=2.5$ and $Q=1/31$, and the
results are shown in Fig. \ref{fig5} for the energy ranges
$-0.8<E<-0.7$ (upper panel) and $-0.4<E<-0.3$ (lower panel)
corresponding to the two respective insulating phases in Fig.
\ref{fig2}. Characteristic dips in the local electron density at the
potential minima are clearly present as marked by the blue arrows in
the lower panel, a signal that the electron density there has
reached maximum below $E=-0.4$. These dips broaden as energy $E$
increases and more sites reach maximum electron density.
Consequently, for Fermi energy $E_F$ between $\pm E_c$, the
real-space electron density after filling all states below $E_F$
receives non-analytic singularities when it drops from maximally
filled to partially filled - similar to the momentum-space electron
density of a metal at the Fermi surface but in the real space
instead. The real-space electron density $\int_{-2t-V}^{E_F}dE
\left|f_n(E)\right|^2$ for all states below the Fermi energy $E_F$
is shown in Fig. \ref{fig9} for $E_F=-0.7$ (upper panel) and
$E_F=-0.3$ (lower panel).

In Sec. \ref{sec:semiclassical}, we have shown that the effective
metal-insulator transition for $V<2t$ is accompanied by a
qualitative change of DOS profile as the influence of the lattice
comes into play. Duality suggested that these should apply equally
to the transitions at $V>2t$. We note that our arguments in Sec.
\ref{sec:semiclassical} near the band edges hold irrespective of
$V$, therefore the band-edge states have the same properties no
matter $V>2t$ or $V<2t$: the eigenstates are effectively and
independently localized within each potential well regardless of the
lattices, therefore the quantized and almost degenerate energy
levels. However, once the electron density reaches maximum at the
potential bottoms, further filling is gradually forced to the
neighboring sites, which occurs sequentially in different potential
wells according to their respective lattice displacement, changing
the spread of the energy eigenvalues and the behavior of DOS.

Note that this transition is unique for $V>2t$. For $V<2t$, the
shallow potential means that the tunneling between the wells becomes
relevant and the gaps in the local integrated DOS between the wells
close before the maximum electron filling is ever reached anywhere.
The transition is then dominated by the metal-insulator transition.
Quite interestingly, the phase transition for $V>2t$ is an analogy
of the metal-insulator transition for $V<2t$ but with the roles of
real space and momentum space interchanged. For $V<2t$ and energy
$\left|E\right|<2t-V$ between the apparent mobility edges, the
system is metallic in the sense that the real-space two-point
Green's function is long-range and the momentum-space electron
density is singular. On the other hand, for $V>2t$ and energy
$\left|E\right|<V-2t$ between the critical points (the ${\rm
insulator}^{*}$ phase in the phase diagram in Fig. \ref{fig2}), the
system has singular real-space density and the electron correlation
is long-range in Fourier space. See Appendix \ref{apd:rsFScor} for
more analytic details. We would like to emphasize that there is no
symmetry breaking in either transition. Rather, they are identified
by the appearance and disappearance of singularities and changes of
physical exponents, yet both are slightly rounded off in the
presence of a finite incommensurate $Q$. Both transitions can be
viewed according to the picture used in Sec. \ref{subsec:2DFS} as a
Lifshitz transition in two dimensions projected to one dimension,
but along different directions. The duality transformation
interchanges the two directions, therefore not surprisingly, it also
exchanges the coordinates in which the transition occurs.

\section{Large $Q$ limit\label{sec:largeQ}}

Up until this point, our discussion has focused on the limit in
which $Q\ll2\pi$ is relatively small. For large $Q/2\pi\sim O(1)$,
the effective magnetic field is large in the corresponding
two-dimensional Hofstadter problem, so the semiclassical theory
ceases to be useful. Conversely, we will consider the case in which
$V/2t$ is small, so in the original 1D formulation of the problem,
we can compute the effects of the incommensurate potential
perturbatively and compare with numerical results. We find
(consistent with our own earlier findings in Ref.
\onlinecite{qoicdw}) that 1) the lowest orders in the perturbation
theory describe Harper's equation satisfactorily so long as $V/2t$
is even moderately small, and 2) all states are de-localized and
form a continuous spectrum separated by the gaps, consistent with
previous expectations.

In the perturbation theory, we begin with a nearest-neighbor
tight-binding model of a one-dimensional chain with energy
dispersion $\epsilon_{k}=2t\cos(k)$, and then consider the gaps
induced by the lowest order scattering from the incommensurate
potential. More specifically, the $n$th order scattering generates
an off-diagonal mass term between momentum $k$ and $k+nQ$, resulting
in a gap of size $\sim V^{n}/W^{n-1}$ centering on energy
$E_n=\epsilon_{k}=\epsilon_{k+nQ}=\pm 2\cos(nQ/2)$. The DOS vanishes
in the gap and the localization length is finite; between the gaps
the spectrum is continuous, and the localization length is infinite.

\begin{figure}
\begin{centering}
\includegraphics[scale=0.3]{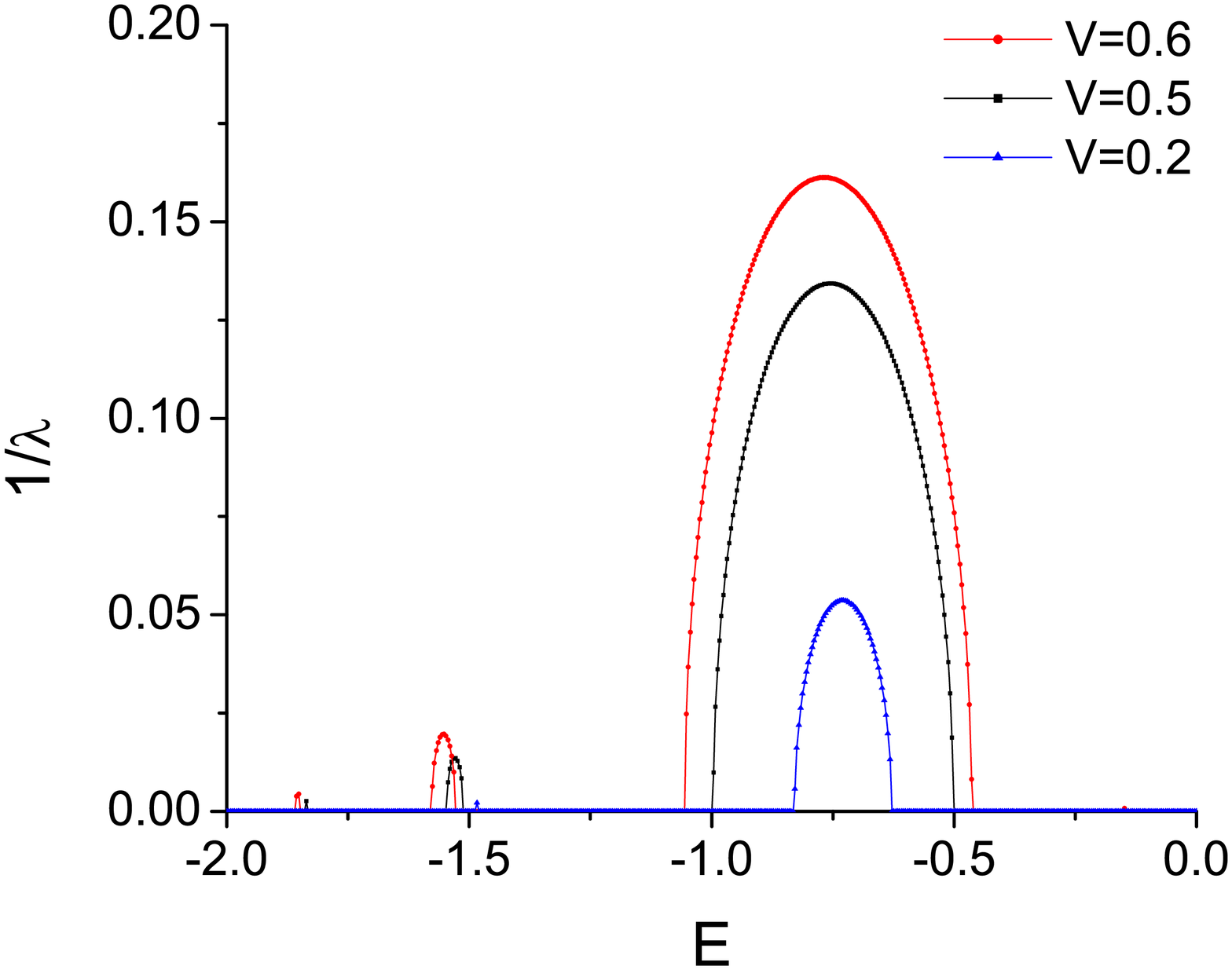}
\includegraphics[scale=0.3]{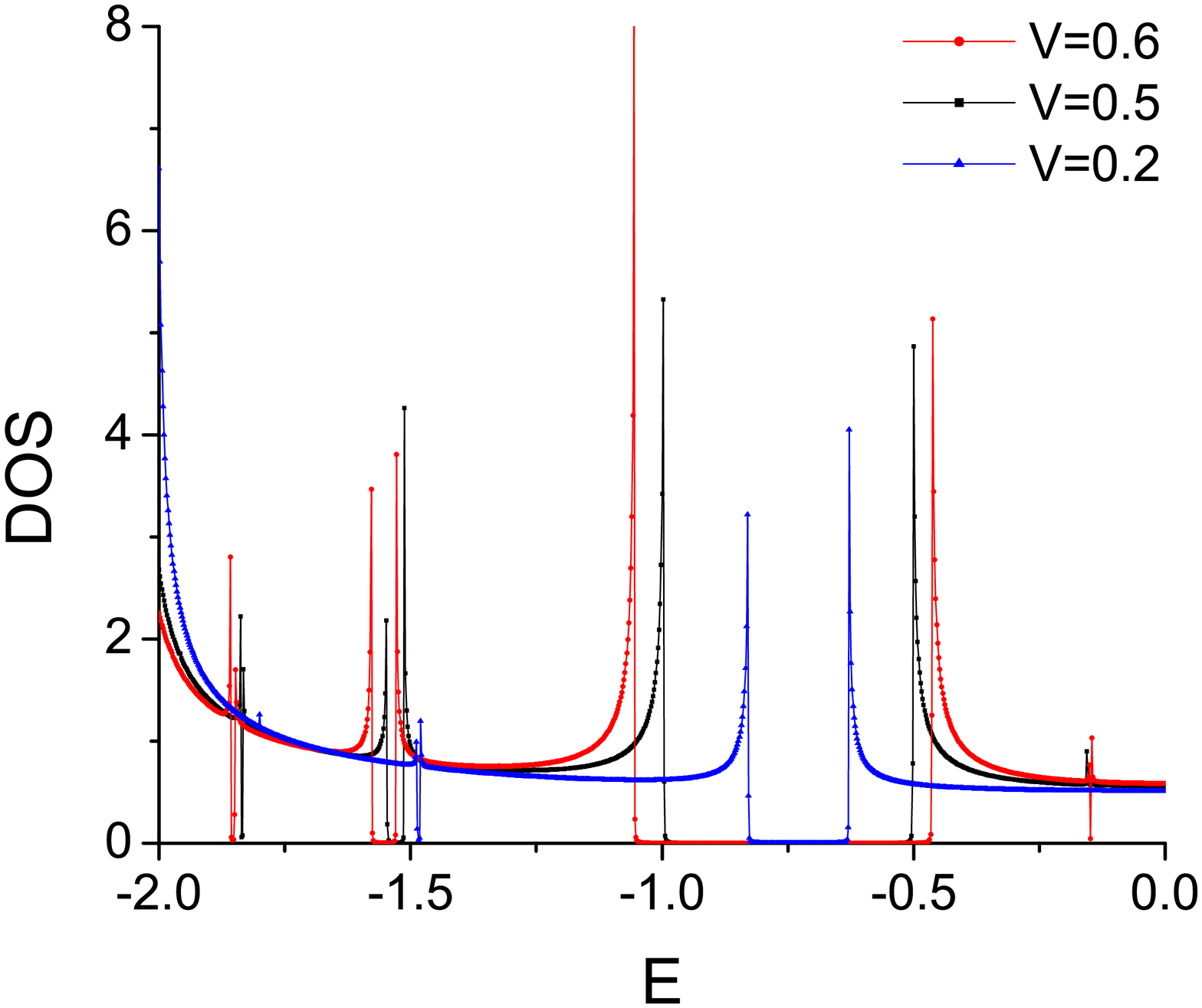}\protect
\protect\caption{The inverse localization length (upper panel) and
the DOS (lower panel) for Harper's equation with $t=1$,
$V=0.2,0.5,0.6$ and a large $Q=2\pi\cdot(\sqrt{5}-1)/2$ over the
energy range $E\in[-2,0]$. Due to particle-hole symmetry, only the
$E<0$ half is shown. The peaks in the DOS correspond to the van Hove
singularities.}\label{fig6}
\par\end{centering}
\end{figure}

To check these conclusions, we numerically compute the inverse
localization length and DOS of Harper's equation with a large $Q$.
One example is shown in Fig. \ref{fig6} with hopping amplitude
$t=1$, potential amplitudes $V=0.2,\ 0.5$, and $0.6$ respectively,
and wave vector (favored in much of the literature)
$Q=2\pi\cdot(\sqrt{5}-1)/2$. A series of gaps are clearly observed,
and the inverse localization length is zero everywhere except inside
the gaps. We note that the centers of the gaps from the first three
orders of perturbation $E_{1}=\pm 2\cos(Q/2)=\pm 0.725$, $E_{2}=\pm
2\cos(Q)=\pm1.475$ and $E_{3}=\pm2\cos(3Q/2)=\pm1.793$ and their
corresponding sizes are remarkably consistent with our numerical
results in Fig. \ref{fig6}. In addition, Van Hove singularities in
the DOS are clearly seen in the DOS on the band edges. These
conclusions are straightforward to generalize to other $Q$.

One major distinction between the incommensurate and commensurate
potentials is that in the former case there is an unending cascade
of higher order gaps while in the latter case new gaps are not
generated beyond the order of perturbation theory corresponding to
the commensurability\cite{qoicdw}. So why can we expect a
finite-order perturbation theory to work for Harper's equation?
Heuristically, the obtained gaps as well as the resulting inverse
localization length are exponentially small ($\sim (V/2t)^n$) which
eventually results in gaps that are unobservable given a finite
energy resolution $\delta$ and system size $L$. In general, for any
given level of desired energy resolution $\delta$ and system size
$L$ there is no essential distinction between the incommensurate
problem and a suitable high order commensurate approximant.

\section{Conclusion and discussions}\label{sec:conclusion}

We have revisited important properties of Harper's equation.  This
problem exhibits many extraordinary features, valid as matters of
principle, if studied with infinite precision, {\it i.e.} assuming
infinite energy resolution, perfect periodicity of the
incommensurate potential, and no higher harmonics to the potential.
This complexity includes conclusions concerning the non-existence of
a true mobility edge which is a consequence of an exact
self-duality, and a self-similar structure of the spectrum. However,
if there is a small parameter, from a physical perspective, in which
there is always a finite limit with which physical quantities can be
controlled,  {\em exponentially small} effects are physically
negligible.

For $Q\ll2\pi$, the previous conclusion that there is a single metal
insulator transition at $V=2t$ for arbitrary energy $E$ holds in the
strict sense. However, in the more physical sense in which a finite
energy resolution is present due to disorder, finite temperature,
system size limit etc., we find two ``effective'' mobility edges at
$\pm E_c=2t-V$ for $V<2t$ and $Q\ll2\pi$; the states near the band
edges are for most physical purposes localized even for a weak
incommensurate potential. Moreover, we have discovered a crossover
at $\pm E_c=V-2t$ for $V>2t$ and $Q\ll2\pi$ that is dual to the
metal insulator transition and characterized by the emergence of a
real-space Fermi surface.

Through both perturbative methods and equivalent modeling, we have
self-consistently re-established the properties of the various
phases of Harper's equation. We note that for small $Q\ll2\pi$, the
two-dimensional Hofstadter problem has a quantized Hall conductance
of $\sigma_{xy}=ne^{2}/h$ if we fill $n$ discrete energy levels near
the band edges, so that $n$ electrons will be transported along the
$\hat{x}$ direction in each cycle of $k_{y}\in[0,2\pi]$. On the
other hand, in the one-dimensional representation of Harper's
equation, $k_{y}$ is the initial phase of the incommensurate
potential, which is adiabatically shifted by exactly one period in
each cycle of $k_{y}$. Since there are $n$ electrons localized in
each period of the potential for the limit we are considering, it is
straightforward to see that $n$ electrons are pumped from one end of
the chain to the other. Such adiabatic pumping process in
quasiperiodic systems has been previously studied in Ref.
\onlinecite{thoulesstransport, qcpumping}. Similar argument also
implies a connection of Harper's equation with certain commensurate
wave vectors to three-dimensional Weyl
semi-metal\cite{dassarma-weyl}.

The methods and conclusions in the main text can be generalized
straightforwardly to models with further neighbor hopping or more
complex potentials\cite{qoicdw}. One interesting example is the
`correlated disorder' of the form $V\cos\left(Qx^\gamma\right)$ with
an incommensurate wave vector $Q$. For $\gamma = 1$, the potential
is reduced to Harper's equation. Based on a different
technique\cite{cordisorder}, mobility edges have been identified for
$\gamma < 1$ and the states are fully localized for $\gamma > 1$.

We acknowledge insightful discussions with Boris Spivak, Pavan
Hosur, Xiao-liang Qi, Andre Broido, Sankar Das Sarma, and Persi
Diaconis.  YZ is supported by the Stanford Institute for Theoretical
Physics, DB is supported by the National Science Foundation Graduate
Research Fellowship under Grant No. DGE-114747, CMJ is supported by
the David and Lucile Packard foundation and National Science
Foundation under Grant No. NSF PHY11-25915, AM is supported by DOE
Office of Basic Energy Sciences under contract No.
DEAC02-76SF00515(AM), and SAK is supported in part by the NSF under
grant DMR-1265593 at Stanford. YZ and CMJ also thanks the
hospitality of KITP where part of this work is completed and support
in part from KITP by the National Science Foundation under Grant No.
NSF PHY11-25915.

\appendix

\section{The accuracy of $\theta$ independence in Harper's equation}\label{apd:thetaind}

In this appendix, we discuss the applicability of physical
quantities' $\theta$ independence. Mathematically, it was proven
that the spectrum of the almost Mathieu equation (Harper's equation)
is independent of the choices of $\theta$ in the thermodynamic limit
and when $Q/2\pi$ is indeed irrational\cite{math1, math2}. Here we
offer a heuristic physical argument.

To start with, we consider an intensive physical quantity
$A_{l}(\theta)$ for a one-dimensional system with finite size
$l\gg1$, where exists in general a sizeable $\theta$ dependence. For
a larger system size $L=nl$, since the interfaces are negligibly
small in comparison with the bulk, we may safely separate the
measurement into various smaller components each with length $l$ but
different phases $\theta_{m}=\theta+(m-1)Ql(\mbox{mod}2\pi)$:
$A_{L=nl}(\theta)=\frac{1}{n}\underset{m}{\sum}A_{l}(\theta_{m})$,
therefore the $\theta$ dependence is averaged out over many
different $\theta_{m}$'s. We may then repeat the procedure for even
larger system sizes, and in each step the $\theta$ dependence scales
down. This suggests that the $\theta$ dependence is inverse
proportionally suppressed as the total system size $L$. Similar
argument works for extensive physical quantities
$A_{L=nl}(\theta)=\underset{m}{\sum}A_{l}(\theta_{m})$, which
receives just an additional factor of the total system size $L$.

It is straightforward to see that this argument is valid for the DOS
$\rho$ (per site), since the electron density (per site) is
extensive (intensive). On the other hand, we show in the following
that the correct self-averaging quantity for the localization is the
inverse localization length $1/\lambda$.

The localization length is defined in terms of the Green's function
$G$ by $G(1,l;\theta)\propto\exp(-l/\lambda_{l}(\theta))$.
Therefore, for a larger system $L=nl$:
\begin{eqnarray}
G(1,L;\theta)&\propto&\underset{m}{\prod}G(ml-l+1,ml;\theta)
\nonumber\\ &\propto&\underset{m}{\prod}G(1,l;\theta_{m})
\propto\exp(-\underset{m}{\sum}l/\lambda_{l}(\theta_{m}))
\end{eqnarray}
Comparing this with
$G(1,L;\theta)\propto\exp(-L/\lambda_{L}(\theta))$, we have on the
order $O(L)$:
\begin{eqnarray}
-L/\lambda_{L}(\theta)\eqsim -\underset{m}{\sum}l/\lambda_{l}(\theta_{m})\nonumber\\
1/\lambda_{L}(\theta)\eqsim\frac{1}{n}\underset{m}{\sum}1/\lambda_{l}(\theta_{m})
\end{eqnarray}

\begin{figure}
\begin{centering}
\includegraphics[scale=0.3]{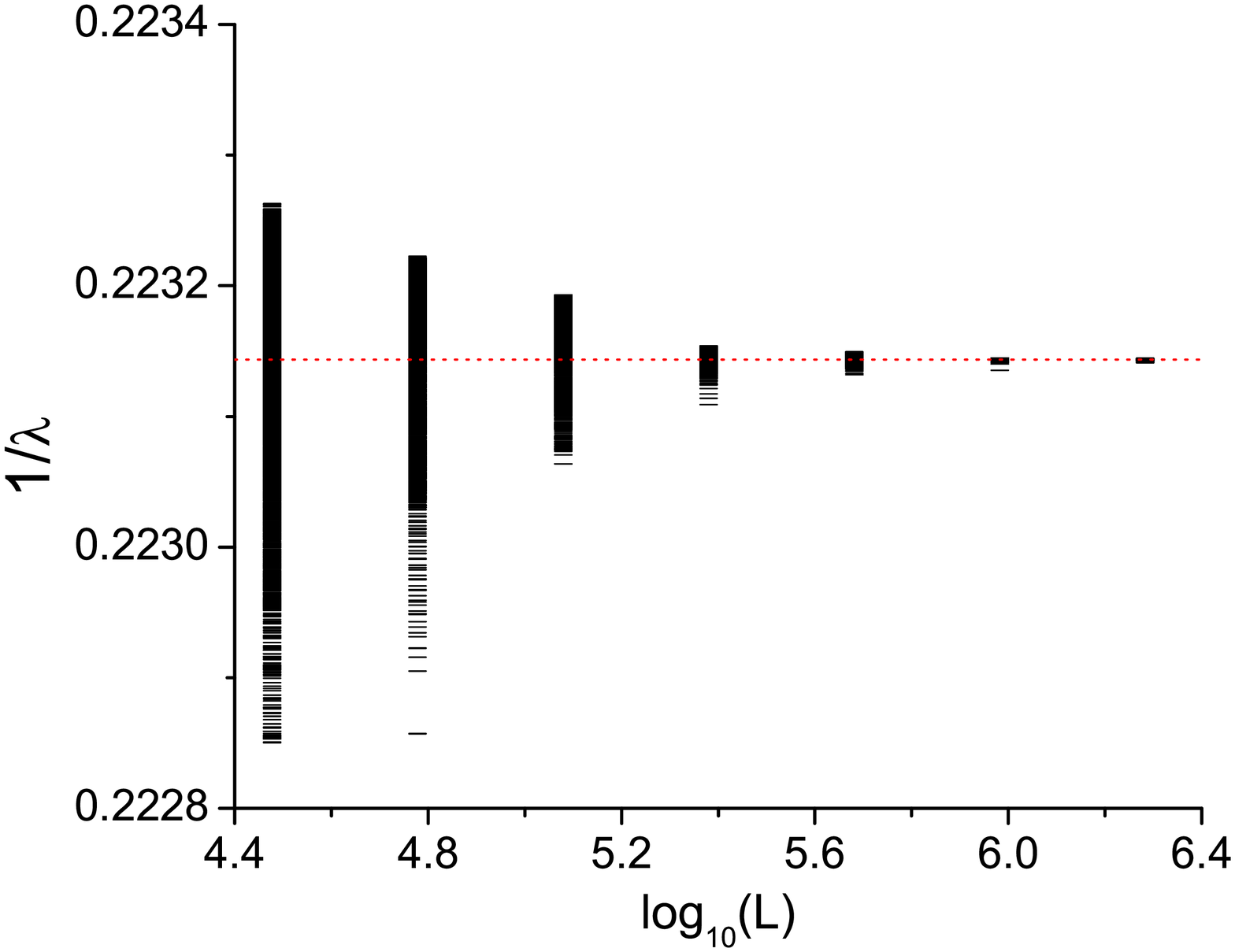}
\includegraphics[scale=0.3]{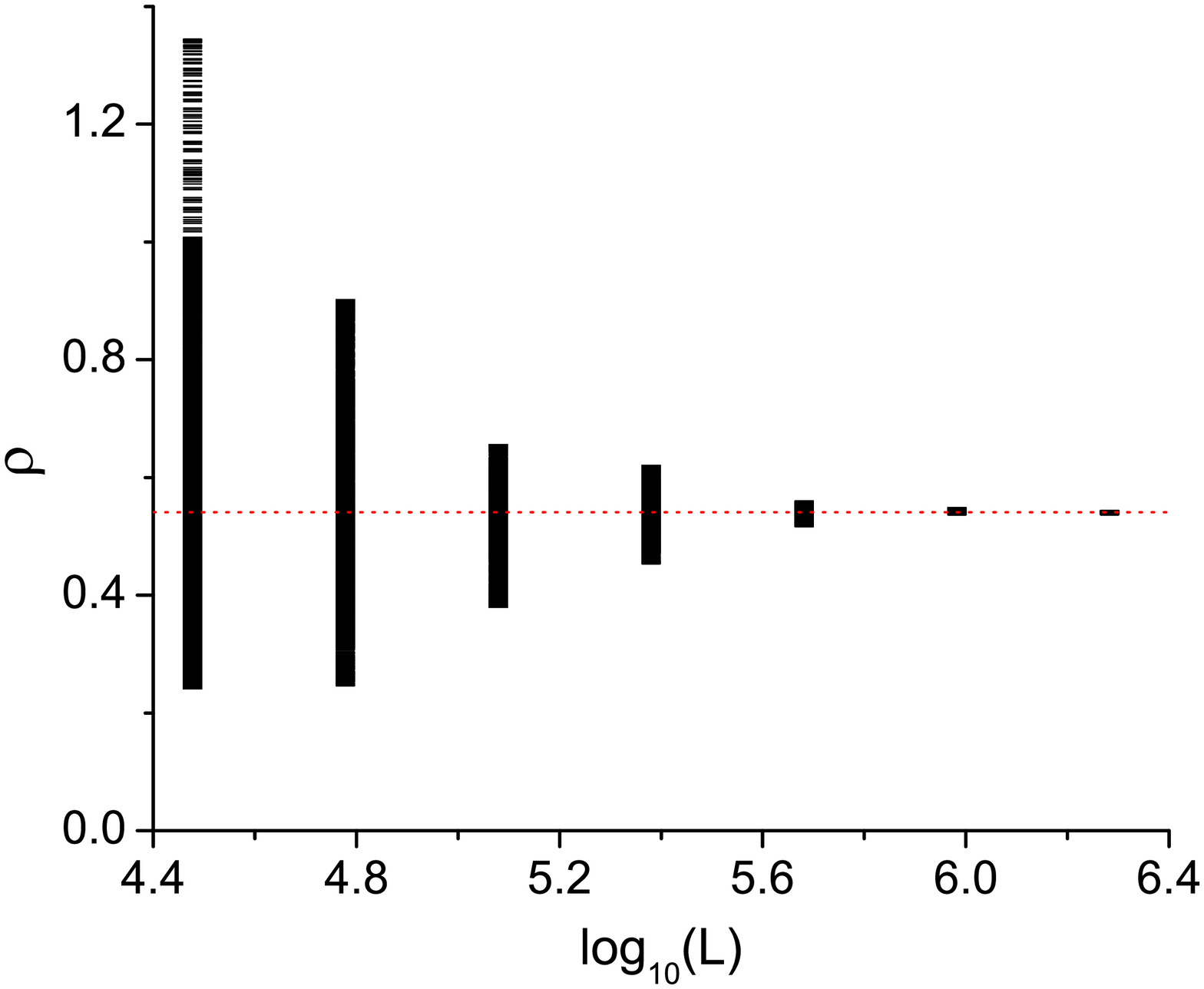}\protect
\protect\caption{The system size effect of the spectrum dependence
on different choices of $\theta$ of the inverse localization length
$1/\lambda$ (upper panel) and the DOS $\rho$ (lower panel) of
Harper's equation with $t=1$, $V=2.5$, $Q=1/31$ and $E=-0.3$. The
red dotted line is the average value over the spectrum for the
largest system size.}\label{fig10}
\par\end{centering}
\end{figure}

We illustrate in Fig. \ref{fig10} the system size effect of the
spectrum dependence of the inverse localization length $1/\lambda$
and the DOS $\rho$ on different choices of $\theta$. For the system
sizes we study in the main text, the $\theta$ dependence has been
suppressed well enough.

\section{Beyond the free-electron approximation near the band
edges}\label{apd:beyondfe}

In Sec. \ref{sec:semiclassical}, we have shown that by expanding the
energy dispersion $\epsilon_k$ to the quadratic order, the
free-electron approximation is fairly consistent with numerical
results on the DOS near the band edges. In this appendix, we show
that the consistency can be made better over a wider range of energy
by taking into account the quartic order terms in the expansion:
\begin{eqnarray}
H&=&-2\cos k - V \cos Qx \nonumber\\&\eqsim& -2 - V + k^2 +
VQ^2x^2/2-\frac{1}{24}\left(2k^4+V Q^4 x^4\right)
\end{eqnarray}
where we use the one-dimensional representation and set $t=1$ and
the origin of the coordinate at the potential minimum for
simplicity.

Let's define the bosonic ladder operators
$a^\dagger\left|n\right\rangle= \sqrt{n+1}\left|n+1\right\rangle$,
$a\left|n\right\rangle= \sqrt{n}\left|n-1\right\rangle$ and
$a^\dagger a\left|n\right\rangle= n\left|n\right\rangle$, with
which:
\begin{eqnarray}
k&=&\frac{i\sqrt \omega}{2} \left(a^\dagger-a\right)\nonumber\\
x&=&\frac{1}{\sqrt\omega}\left(a^\dagger+a\right)
\end{eqnarray}
the canonical commutation relations are preserved. Here $\omega =
Q\sqrt{2V}$.

As a consistency check, we first consider the quadratic order:
\begin{eqnarray}
H^{(2)}&=&k^2+\omega^2 x^2/4=\frac{\omega
}{4}\left[\left(a^\dagger+a\right)^2-\left(a^\dagger-a\right)^2\right]\nonumber\\&=&\frac{\omega}{2}\left(a^\dagger
a+a a^\dagger\right)=\omega\left(n+1/2\right)
\end{eqnarray}
which is diagonal in the $\left|n\right\rangle$ representation and
has equally spaced energy levels. This is the harmonic oscillator.

We can generalize the expression in terms of the ladder operators to
the quartic order:
\begin{eqnarray}
H^{(4)}&=&-\frac{1}{24}\left(2k^4+V Q^4 x^4\right) \nonumber\\
&=&-\frac{\omega^2}{96}\left[\frac{1}{2}\left(a^\dagger
-a\right)^4+\frac{1}{V}\left(a^\dagger +a\right)^4\right] \nonumber
\\
&=&-\frac{\omega^2}{96}\left[\left(\frac{1}{2}+\frac{1}{V}\right)a^\dagger
a^\dagger a^\dagger a^\dagger
-\left(4n+6\right)\left(\frac{1}{2}-\frac{1}{V}\right)a^\dagger
a^\dagger+ \mbox{H.c.}\nonumber\right.\\
&+& \left.\left(6n^2+6n+3\right)
\left(\frac{1}{2}+\frac{1}{V}\right)\right]
\end{eqnarray}
which now has off-diagonal components. Still, for the lowest energy
levels we can get a sound approximation with a reasonable up-limit
for $n$ and exact diagonalize the Hamiltonian in the
$\left|n\right\rangle$ representation. The results for $V=1.6$ and
$Q=1/31$ are shown in the blue marks in Fig. \ref{fig4}.

It is straightforward to generalize to higher order expansions.

\section{The tunneling amplitude from WKB approximation}\label{apd:functionT}

To make a more quantitative estimate of the tunneling amplitude
between the neighboring wells $t_{\rm eff}$ in a slowly varying
potential $V\cos(Qx)$, we generalize the canonical WKB approximation
to the current dispersion relation:
\begin{eqnarray}
2t\cos\left[k\left(x\right)\right]&-&V\cos\left(Qx\right)=E
\nonumber
\\
t\left[e^{\kappa\left(x\right)}+e^{-\kappa\left(x\right)}\right]&=&-V\cos
\left(Qx\right)-E
\\
e^{\pm\kappa(x)}&=& -E/2t-\left(V/2t\right)\cos
\left(Qx\right)\nonumber
\\
&\pm&\sqrt{\left[E/2t+\left(V/2t\right)\cos
\left(Qx\right)\right]^{2}-1} \nonumber
\end{eqnarray}
where the energy $E$ is near the lower band edge and
$x\in[0,2\pi/Q]$ labels the sites between the neighboring wells.
Inside the barrier, the wavefunction decays as $
e^{-\kappa\left(x\right)}$ where $\kappa=ik$ is the imaginary
momentum. The tunneling amplitude across all sites among the barrier
is
\begin{eqnarray}
t_{\rm eff} &\propto&\underset{x}{\prod}e^{-\kappa(x)}  \propto \exp
 -\underset{x}{\sum} \log \left\{ -E/2t\right.
\\
&-&\left(V/2t\right)\cos
\left(Qx\right)+\left.\sqrt{\left[E/2t+\left(V/2t\right)\cos
\left(Qx\right)\right]^{2}-1}\right\}\nonumber
\end{eqnarray}

Replacing the summation with the integral $\int_{0}^{2\pi/Q} dx$, it
is straightforward to see that $t_{\rm eff} \propto \exp\left[
-(2\pi\alpha/Q)\right]$ is exponentially suppressed by the distance
$2\pi/Q$ between the potential wells, where
\begin{eqnarray}
\alpha&=&\frac{1}{2\pi} \int_{0}^{2\pi} \log \left\{
\sqrt{\left[E/2t+\left(V/2t\right)\cos
\left(y\right)\right]^{2}-1}\right. \nonumber\\
& & \left. -E/2t-\left(V/2t\right)\cos \left(y\right)\right\} dy
\end{eqnarray}
is an increasing function of $V/2t$, since a higher barrier tends to
reduce the tunneling amplitude. At the band bottom $E=-2t-V$, the
dependence of $t_{\rm eff}$ on $V/2t$ and $Q$ is qualitatively
consistent with Eq. \ref{eq:teff} from simple variational analysis.

\section{Application to metal-insulator transition of generalized Harper's equation}\label{apd:gaa}

It is straightforward to generalize Harper's equation to include
further neighbor hoppings and higher harmonic components in the
incommensurate potential:

\begin{eqnarray}
t\underset{n'}{\sum}a_{n'-n}f_{n'}+Vf{}_{n}\underset{m}{\sum}b_{m}\exp\left(iQm\cdot
n\right)=Ef_{n} \label{eq:gaa}\end{eqnarray} where $a_{n}$ and
$b_{m}$ are given parameters and $Q$ is an incommensurate wave
vector. In this appendix, we briefly discuss locating the
metal-insulator transition separating the insulating phase for large
$V$ and metallic phase for small $V$ in the $Q/2\pi\ll1$ limit from
a two-dimensional Hofstadter problem perspective.

The generalized Harper's equation in Eq. \ref{eq:gaa} is equivalent
to a two-dimensional lattice problem with dispersion
$\epsilon(\vec{k})=t\underset{n}{\sum}a_{n}\exp\left(ik_{x}n\right)+V\underset{m}{\sum}b_{m}\exp\left(ik_{y}m\right)$
in a magnetic field $B_{z}=Q/2\pi$. According to Sec.
\ref{sec:semiclassical}, the states at energy $E$ are fully
de-localized (localized) if the Fermi surface defined by
$\epsilon(\vec{k})=E$ is open along the $\hat{y}$ direction
($\hat{x}$ direction). Yet, as we have shown in Sec.
\ref{sec:paradox}, the \emph{true} metal-insulator transition is
determined by the quantum tunnelings across the $\hat{x}$ and
$\hat{y}$ directions when the Fermi surface is closed:
\begin{eqnarray}
\tilde{t}_{x}=\exp\left(-\frac{1}{2\pi}\intop_{0}^{2\pi}\left|Imk_{y}(k_{x})\right|dk_{x}\right)
\nonumber \\
\tilde{t}_{y}=\exp\left(-\frac{1}{2\pi}\intop_{0}^{2\pi}\left|Imk_{x}(k_{y})\right|dk_{y}\right)
\label{eq:txtyeff}
\end{eqnarray}
where $k_{x}(k_{y})$ and $k_{y}(k_{x})$ are solutions of the
dispersion $\epsilon(\vec{k})=E$ for a given $k_{y}$ and $k_{x}$,
respectively. It is also straightforward to see that $\tilde{t}_{y}$
and $\tilde{t}_{x}$ are related to the $1/Q$ coefficients in the
one-dimensional tunneling amplitude between the neighboring
potential wells in the original Harper's equation and dual Harper's
equation, respectively. We have listed some examples as follows:

\emph{Generalized Harper's equations with implicit self-duality}: In
Ref. \onlinecite{gaa}, mobility edges have been identified for the
following generalized Harper's equations:
\begin{eqnarray}
t\left(u_{n-1}+u_{n+1}\right)+V_{n}u_{n}=Eu_{n}
\nonumber\\
V_{n}=2V\frac{\cos\left(Qn+\phi\right)}{1-r\cos\left(Qn+\phi\right)}
\end{eqnarray}
which can be mapped to an equivalent non-interacting two-dimensional
lattice model in the presence of an incommensurate magnetic field of
$B_{z}=Q/2\pi$ with dispersion:
\begin{eqnarray}
2t\cos k_{x}+2V\frac{\cos k_{y}}{1-r\cos k_{y}}&=&E \\
2t\cos k_{x}\left(1-r\cos k_{y}\right)+2V\cos
k_{y}&=&E\left(1-r\cos k_{y}\right) \nonumber\\
-2r t\cos k_{x}\cos k_{y}+2t\cos k_{x}&+&\left(2V+r E\right)\cos
k_{y}=E \nonumber
\end{eqnarray}

It is straightforward to see that for $2V+r E>2t$ ($2V+r E<2t$) the
constant energy contour at $E$ is elongated along the $\hat{x}$
direction ($\hat{y}$ direction), $\tilde{t}_{x}$ ($\tilde{t}_{y}$ )
is dominant and the model is localized (de-localized) along the
initial $\hat{x}$ direction. Clearly, there is a mobility edge at
$E=2\left(t-V\right)/r$, where the Fermi surface is symmetric under
$k_{x}\leftrightarrow k_{y}$ and $\tilde{t}_{x}=\tilde{t}_{y}$. This
result is fully consistent with the conclusions in Ref.
\onlinecite{gaa}. It is also simple to check the consistency of our
argument with the other model in Ref. \onlinecite{gaa}:
\begin{eqnarray}
V_{n}=2V\left[1-\cos\left(Qn+\phi\right)\right]/\left[1+r\cos\left(Qn+\phi\right)\right]
\end{eqnarray}

Similar self-duality and mobility edge arguments can be generalized
to two-dimensional incommensurate Hofstadter problems with
dispersion of the following form:
\begin{eqnarray}
h\left(\cos k_{x}\right)+\frac{a+b\cdot h\left(\cos
k_{y}\right)}{c+h\left(\cos k_{y}\right)}=E
\end{eqnarray}
where $h(t)$ is a power series of $t$ and $a$, $b$, $c$ are
parameters. The mobility edge is at $E=b-c$.

\begin{figure}
\begin{centering}
\includegraphics[scale=0.4]{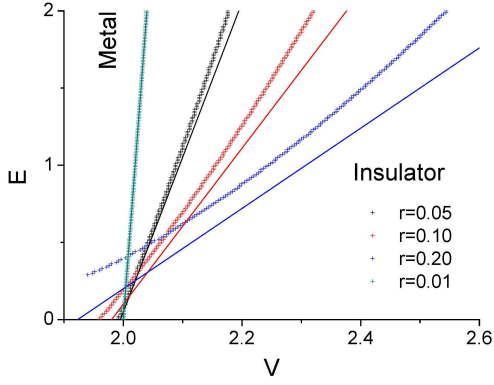}
\protect\caption{Phase diagram and mobility edge of generalized
Harper's equation in Eq. \ref{apd:nnnharper} for various
next-nearest and nearest neighbor hopping amplitude ratio $r$. The
results are obtained through calculations of Eq. \ref{eq:txtyeff}.
The solid lines are the exact mobility edges\cite{gaa} for the
approximants with hopping amplitudes $t_n=r^{n-1} t$.}\label{figa3}
\par\end{centering}
\end{figure}

\emph{Generalized Harper's equation with nearest and next-nearest
neighbor hopping:} we don't always have the luxury of symmetry for
the determination of the relative strength of $\tilde{t}_{y}$ and
$\tilde{t}_{x}$ in a generic dispersion $\epsilon(\vec{k})$,
therefore, we need to calculate $\tilde{t}_{y}$ and $\tilde{t}_{x}$
using Eq. \ref{eq:txtyeff}. One simple example is:
\begin{eqnarray}
t\left[\left(u_{n-1}+u_{n+1}\right)+r\left(u_{n-2}+u_{n+2}\right)\right]+V\cos(Qn)u_{n}=Eu_{n}\label{apd:nnnharper}
\end{eqnarray}
where $r$ is the ratio of the next-nearest neighbor to the nearest
neighbor hopping amplitude. We numerically calculate $\tilde{t}_{y}$
and $\tilde{t}_{x}$ from the dispersion relation of the
corresponding two-dimensional Hofstadter problem
$\epsilon(\vec{k})=t\left(\cos k_{x}+r\cos2k_{x}\right)+V\cos
k_{y}$, and the metal-insulator phase boundaries for various $r$ is
shown in Fig. \ref{figa3}. In addition, the exact locations of the
mobility edges\cite{gaa} for approximants of Eq. \ref{apd:nnnharper}
with hopping amplitudes $t_n=r^{n-1} t$, $n=1,2,\cdots$ are shown
for comparative purposes.

\section{Characteristic correlations of a real-space Fermi
surface}\label{apd:rsFScor}

From the duality transformation in Eq. \ref{eq:duality} we obtain:
\begin{eqnarray}
\left\langle c_{n}^{\dagger}c_{0}\right\rangle
=\frac{1}{L}\underset{m,m'=1}{\sum^L}\tilde{c}_{m}^{\dagger}\tilde{c}_{m'}\exp\left(imnQ+i\theta
m-i\theta m'\right)
\end{eqnarray}
where $\tilde{c}$ is the electron operator for the dual Harper's
equation with $V>2t$. We take advantage of the fact that for a
quasiperiodic system the correlations are independent of the choices
of $\theta$:
\begin{eqnarray}
\left\langle c_{n}^{\dagger}c_{0}\right\rangle &=&\frac{1}{2\pi
L}\underset{m,m'=1}{\sum^L}\tilde{c}_{m}^{\dagger}\tilde{c}_{m'}\int
d\theta \exp\left(imnQ+i\theta
m-i\theta m'\right) \nonumber\\
&=&\frac{1}{L}\underset{m,m'}{\sum}\tilde{c}_{m}^{\dagger}\tilde{c}_{m'}\exp\left(imnQ\right)\delta_{mm'}\nonumber\\
&=&\frac{1}{L}\underset{m}{\sum}\tilde{c}_{m}^{\dagger}\tilde{c}_{m}\exp\left(imnQ\right)=\tilde\rho\left(nQ\right)
\label{eq:cordual}
\end{eqnarray}
where $\tilde \rho(nQ)$ is the Fourier transform of the electron
density of the dual model.

In Sec. \ref{sec:semiclassical}, we have shown that for Harper's
equation with $V<2t$ and $Q\ll2\pi$, the electron states near the
band edge are effectively localized in the physical limit with
finite energy resolution $\delta$, and the Green's function
$\left\langle c_{n}^{\dagger}c_{0}\right\rangle $ is exponentially
suppressed as a function of $n$; otherwise, near the band center
between the mobility edges, $\left\langle
c_{n}^{\dagger}c_{0}\right\rangle$ is finite for all $n$. According
to Eq. \ref{eq:cordual}, for the dual Harper's equation with $V>2t$,
the Fourier transform of the density $\tilde\rho\left(nQ\right)$
near the band edge is order $\sim O(1)$ when $n$ is small, since the
electron density has an induced charge density wave with wave vector
$Q$; however, at large $n\rightarrow \infty$,
$\tilde\rho\left(nQ\right)\rightarrow 0$ is suppressed
exponentially. In contrast, around the band center within the energy
range $E\in(2t-V,V-2t)$, the emergence of real-space electron
density singularities (see our conclusions in Sec.
\ref{sec:hiddenPT}) allows the Fourier components $\rho(nQ)$ to be
finite for all $n\in\mathbb{Z}$.

\end{document}